\pdfoutput=1

\documentclass[
  reprint,
  superscriptaddress,
 amsmath,amssymb,
 aps,prb
]{revtex4-1}

\usepackage{graphicx}
\usepackage{dcolumn}
\usepackage{bm}
\usepackage{color}

\bmdefine{\bdi}{i}
\bmdefine{\bdj}{j}
\bmdefine{\bdx}{x}
\bmdefine{\bdy}{y}
\bmdefine{\bdr}{r}
\bmdefine{\bdR}{R}
\bmdefine{\bdS}{S}
\bmdefine{\bdL}{L}
\bmdefine{\bdJ}{J}
\bmdefine{\bdA}{A}
\bmdefine{\bdE}{E}
\bmdefine{\bdD}{D}
\bmdefine{\bdQ}{Q}
\bmdefine{\bdq}{q}
\bmdefine{\bdzero}{0}
\bmdefine{\bdv}{v}
\bmdefine{\bde}{e}
\bmdefine{\bddelta}{\delta}
\bmdefine{\bdnabla}{\nabla}
\bmdefine{\bda}{a}
\bmdefine{\bdb}{b}

\begin{document}

\title{
  Interband magnon drag in ferrimagnetic insulators
}

\author{Naoya Arakawa}
\email{arakawa@phys.chuo-u.ac.jp}
\affiliation{The Institute of Science and Engineering,
  Chuo University, Bunkyo, Tokyo, 112-8551, Japan}

\date{\today}


\begin{abstract}
  We propose a new drag phenomenon, an interband magnon drag,
  and report on interaction effects and multiband effects
  in magnon transport of ferrimagnetic insulators. 
  We study
  a spin-Seebeck coefficient $S_{\textrm{m}}$,
  a magnon conductivity $\sigma_{\textrm{m}}$,
  and a magnon thermal conductivity $\kappa_{\textrm{m}}$
  of interacting magnons for a minimal model of ferrimagnetic insulators
  using a $1/S$ expansion of the Holstein-Primakoff method,
  the linear-response theory, and a method of Green's functions.
  We show that the interband magnon drag
  enhances $\sigma_{\textrm{m}}$ and reduces $\kappa_{\textrm{m}}$,
  whereas its total effects on $S_{\textrm{m}}$ are small.
  This drag results from the interband momentum transfer
    induced by the magnon-magnon interactions.
  We also show that
  the higher-energy band magnons contribute to $S_{\textrm{m}}$, $\sigma_{\textrm{m}}$,
  and $\kappa_{\textrm{m}}$
  even for temperatures smaller than the energy difference between the two bands.

\end{abstract}
\maketitle


\section{Introduction}

Magnon transport is the key to
understanding spintronics and spin-caloritronics phenomena of
magnetic insulators~\cite{Saitoh-Nature,Saitoh-NatMat,Bauer-review}.
For example, a magnon spin current is vital
for the spin Seebeck effect~\cite{Saitoh-NatMat,PM-SSE,AF-SSE-theory,AF-SSE1,AF-SSE2}.
Magnon transport is important also for
other relevant phenomena~\cite{SMR,SpinPert,ThCond-YIG,Nonlocal,MagChem-Bauer,MagSpinCond}. 

There are two key issues about magnon transport in ferrimagnetic insulators. 
One is about multiband effects.
Yttrium iron garnet (YIG) is a ferrimagnetic insulator
used in various spintronics or spin-caloritronics
phenomena~\cite{Saitoh-Nature,Saitoh-NatMat,Bauer-review,SMR,SpinPert,ThCond-YIG,Nonlocal,MagChem-Bauer}.
Its magnons have been often approximated as 
those of a ferromagnet.
However,
a study using its realistic model~\cite{Bauer-PRL} showed that
not only the lowest-energy band magnons,
which could be approximated as those of a ferromagnet, 
but also the second-lowest-energy band magnons
should be considered except for sufficiently low temperatures.
Since the experiments using YIG
are performed typically
at room temperature~\cite{Saitoh-Nature,Saitoh-NatMat,Bauer-review,SMR,SpinPert,Nonlocal,MagChem-Bauer},
it is necessary to clarify the effects of 
the higher-energy band magnons on the magnon transport.
The other is about interaction effects. 
The magnon-magnon interactions are usually neglected.
However,
their effects may be drastic in a ferrimagnet
because 
they can induce the interband momentum transfer,
which 
is expected to cause an interband magnon drag
by analogy with various drag
phenomena~\cite{CD-exp1,CD-exp2,CD-exp3,CD-theory1,CD-theory2,PD-theory1,PD-theory2,PD-theory3,Ogata,vdW-drag,MD-theory1,MD-theory2,MD-theory3,MD-exp1,MD-exp2,MD-exp3,MD-exp4,SCD-theory,NA-SCD,SCD-exp,ColdAtom1,ColdAtom2,ColdAtom3,NonlocDrag,EnergyDrag,PhotonDrag}.
Nevertheless,
it remains unclear
how the magnon-magnon interactions affect the magnon transport.

In this paper,
we provide the first step towards resolving the above issues 
and propose a new drag phenomenon, the interband magnon drag.
We derive three transport coefficients
of interacting magnons 
for a two-sublattice ferrimagnet 
and numerically evaluate their temperature dependences.
We show that
the interband magnon drag enhances a magnon conductivity
and reduces a magnon thermal conductivity,
whereas its total effects on a spin-Seebeck coefficient are small.
We also show that
the higher-energy band magnons
contribute to these transport coefficients
even for temperatures lower than the energy splitting of the two bands. 

\begin{figure}
  \includegraphics[width=54mm]{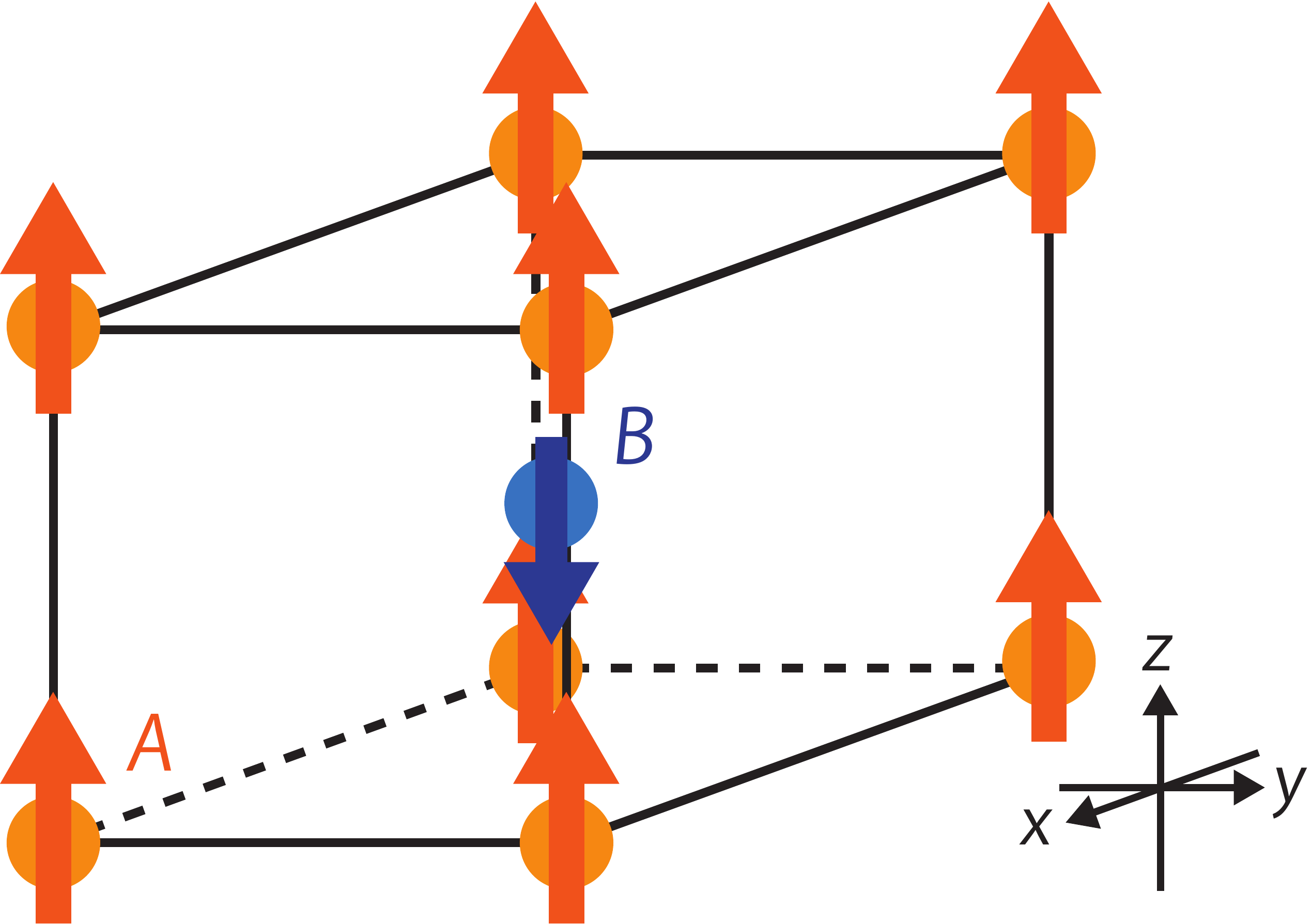}
  \caption{\label{fig1}
    Our ferrimagnetic insulator.
    The up or down arrows represent the spins on the $A$ or $B$ sublattice,
    respectively.
    The $x$, $y$, and $z$ axes are also shown. 
  }
\end{figure}

\section{Model}

Our ferrimagnetic insulator is described by
\begin{align}
  H=2J\sum_{\langle i,j\rangle}\bdS_{i}\cdot\bdS_{j}
  -h\sum_{i=1}^{N/2}S_{i}^{z}-h\sum_{j=1}^{N/2}S_{j}^{z},\label{eq:Hspin}
\end{align}
where the first term is 
the Heisenberg exchange interaction between nearest-neighbor spins,
and the others are the Zeeman energy of a weak magnetic field ($|h|\ll J$).
(The ground-state magnetization is aligned parallel to the magnetic field.)
We have disregarded the dipolar interaction and the magnetic anisotropy,
which are usually much smaller than $J$~\cite{Bauer-PRL,YIG-1stPrinc}.
For concreteness, 
we consider a two-sublattice ferrimagnet
on the body-centered cubic lattice (Fig. \ref{fig1}); 
$i$'s and $j$'s in Eq. (\ref{eq:Hspin})
are site indices of the $A$ and $B$ sublattice, respectively.
There are $N/2$ sites per sublattice.
Our model can be regarded as
a minimal model of ferrimagnetic insulators 
because a ferrimagnetic state, the spin alignments of which are given by
$\bdS_{i}={}^{t}(0\ 0\ S_{A})$ for all $i$'s 
and $\bdS_{j}={}^{t}(0\ 0\ -S_{B})$ for all $j$'s,
is stabilized for $J>0$ with the weak magnetic field.
We set $\hbar=1$, $k_{\textrm{B}}=1$, and $a=1$, where
$a$ is the lattice constant. 

To describe magnons of our ferrimagnetic insulator,
we rewrite Eq. (\ref{eq:Hspin}) by using the Holstein-Primakoff method~\cite{HP}.
By applying the Holstein-Primakoff transformation~\cite{Nakamura,NA-PRL,NA-JPSJ}
to Eq. (\ref{eq:Hspin})
and using a $1/S$ expansion~\cite{Nakamura,Oguchi,NA-PRL}
and the Fourier transformation of magnon operators, 
we can write Eq. (\ref{eq:Hspin}) in the form
\begin{align}
  H=H_{\textrm{KE}}+H_{\textrm{int}}.
\end{align}
Here $H_{\textrm{KE}}$ represents the kinetic energy of magnons, 
\begin{align}
  H_{\textrm{KE}}=
  \sum_{\bdq}
  \left(a_{\bdq}^{\dagger}\ b_{\bdq}\right)
  \left(
  \begin{array}{@{\,}cc@{\,}}
    \epsilon_{AA} & \epsilon_{AB}(\bdq)\\[3pt]
    \epsilon_{AB}(\bdq) & \epsilon_{BB} 
  \end{array}
  \right)
  \left(
  \begin{array}{@{\,}c@{\,}}
    a_{\bdq}\\[3pt]
    b_{\bdq}^{\dagger}
  \end{array}
  \right),\label{eq:HKE}  
\end{align}
where $\epsilon_{AA}=2J_{\bdzero}S_{B}+h$, $\epsilon_{AB}(\bdq)=2\sqrt{S_{A}S_{B}}J_{\bdq}$,
$\epsilon_{BB}=2J_{\bdzero}S_{A}-h$,
and $J_{\bdq}=8J\cos\frac{q_{x}}{2}\cos\frac{q_{y}}{2}\cos\frac{q_{z}}{2}$;
$H_{\textrm{int}}$ represents the leading terms of magnon-magnon interactions,
\begin{align}
  &H_{\textrm{int}}=-\frac{1}{N}
  \sum_{\bdq_{1},\bdq_{2},\bdq_{2},\bdq_{4}}\delta_{\bdq_{1}+\bdq_{2},\bdq_{3}+\bdq_{4}}
  (2J_{\bdq_{1}-\bdq_{3}}a_{\bdq_{1}}^{\dagger}a_{\bdq_{3}}b_{\bdq_{4}}^{\dagger}b_{\bdq_{2}}\notag\\
  &+\sqrt{\frac{S_{A}}{S_{B}}}J_{\bdq_{1}}a_{\bdq_{1}}b_{\bdq_{2}}^{\dagger}b_{\bdq_{3}}b_{\bdq_{4}}
  +\sqrt{\frac{S_{B}}{S_{A}}}J_{\bdq_{1}}b_{\bdq_{1}}a_{\bdq_{2}}^{\dagger}a_{\bdq_{3}}a_{\bdq_{4}})
  +(\textrm{H.c.}).\label{eq:Hint}
\end{align}
We can also express $H_{\textrm{KE}}$ as a two-band Hamiltonian
by using the Bogoliubov transformation~\cite{Nakamura,NA-PRL,NA-JPSJ}:
\begin{align}
  H_{\textrm{KE}}=\sum_{\bdq}[\epsilon_{\alpha}(\bdq)\alpha_{\bdq}^{\dagger}\alpha_{\bdq}
    +\epsilon_{\beta}(\bdq)\beta_{\bdq}\beta_{\bdq}^{\dagger}],
\end{align}
where $\epsilon_{\alpha}(\bdq)=h+J_{\bdzero}(S_{B}-S_{A})+\Delta\epsilon_{\bdq}$,
$\epsilon_{\beta}(\bdq)=-h+J_{\bdzero}(S_{A}-S_{B})+\Delta\epsilon_{\bdq}$,
and $\Delta\epsilon_{\bdq}=\sqrt{J_{\bdzero}^{2}(S_{A}+S_{B})^{2}-4S_{A}S_{B}J_{\bdq}^{2}}$.
For $S_{A}>S_{B}$ we have $\epsilon_{\alpha}(\bdq)<\epsilon_{\beta}(\bdq)$. 
Note that the Bogoliubov transformation is given by
$a_{\bdq}=(U_{\bdq})_{A\alpha}\alpha_{\bdq}+(U_{\bdq})_{A\beta}\beta_{\bdq}^{\dagger}$
and $b_{\bdq}^{\dagger}=(U_{\bdq})_{B\alpha}\alpha_{\bdq}+(U_{\bdq})_{B\beta}\beta_{\bdq}^{\dagger}$,
where $(U_{\bdq})_{A\alpha}=(U_{\bdq})_{B\beta}=\cosh\theta_{\bdq}$,
$(U_{\bdq})_{A\beta}=(U_{\bdq})_{B\alpha}=-\sinh\theta_{\bdq}$,
and these hyperbolic functions satisfy
$\cosh2\theta_{\bdq}=[J_{\bdzero}(S_{A}+S_{B})]/\Delta\epsilon_{\bdq}$
and
$\sinh2\theta_{\bdq}=(2\sqrt{S_{A}S_{B}}J_{\bdq})/\Delta\epsilon_{\bdq}$.
Then,
by using the Bogoliubov transformation,
we can decompose $H_{\textrm{int}}$ into the intraband and the interband components~\cite{NA-AF}.
Because of these properties,
our model is a minimal model to study the two key issues explained above. 

\section{Derivations of transport coefficients}

We consider
three transport coefficients:
a spin-Seebeck coefficient $S_{\textrm{m}}$,
a magnon conductivity $\sigma_{\textrm{m}}$,
and a magnon thermal conductivity $\kappa_{\textrm{m}}$.
They are given by 
$S_{\textrm{m}}=L_{12}$, $\sigma_{\textrm{m}}=L_{11}$, and $\kappa_{\textrm{m}}=L_{22}$, 
where $L_{\mu\eta}$'s are defined as
\begin{align}
  \bdj_{S}=L_{11}\bdE_{S}+L_{12}\Bigl(-\frac{\nabla T}{T}\Bigr),\\
  \bdj_{Q}=L_{21}\bdE_{S}+L_{22}\Bigl(-\frac{\nabla T}{T}\Bigr).
\end{align}
Here
$\bdj_{S}$ and $\bdj_{Q}$ are
magnon spin and heat, respectively, current densities,
$\bdE_{S}$ is a nonthermal external field,
and $\nabla T$ is a temperature gradient.
(Note that
one of the possible choices of $\bdE_{S}$ is a magnetic-field gradient~\cite{Nakata}.)
$L_{21}=L_{12}$ holds owing to the Onsager reciprocal theorem.
It should be noted that
although $\kappa_{\textrm{m}}$ is generally given by
$\kappa_{\textrm{m}}=L_{22}-\frac{L_{21}L_{12}}{L_{11}}$, 
our definition $\kappa_{\textrm{m}}=L_{22}$
is sufficient to describe the thermal magnon transport at low temperatures
at which the magnon picture is valid
because the $L_{22}$ gives the leading temperature dependence.
Since a magnon chemical potential is zero in equilibrium,
$\bdj_{Q}=\bdj_{E}$,
where $\bdj_{E}$ is a magnon energy current density.
Hereafter we focus on the magnon transport
with $\bdE_{S}$ or $(-\nabla T/T)$ applied along the $x$ axis. 

We express $L_{\mu\eta}$'s 
in terms of the correlation functions 
using the linear-response theory~\cite{Kubo,Luttinger,Streda,Ogata,AGD,Eliashberg,Kontani}. 
First,
$L_{12}$ is given by
\begin{align}
  L_{12}=\lim_{\omega\rightarrow 0}
  \frac{\Phi^{\textrm{R}}_{12}(\omega)-\Phi^{\textrm{R}}_{12}(0)}{i\omega},\label{eq:L12}
\end{align}
where 
$\Phi^{\textrm{R}}_{12}(\omega)=\Phi_{12}(i\Omega_{n}\rightarrow \omega+i\delta)$ ($\delta=0+$),
\begin{align}
  \Phi_{12}(i\Omega_{n})=\int_{0}^{T^{-1}}d\tau e^{i\Omega_{n}\tau}\frac{1}{N}
  \langle T_{\tau}J_{S}^{x}(\tau)J_{E}^{x}\rangle,\label{eq:Phi12}
\end{align}
and $\Omega_{n}=2\pi T n$ ($n>0$). 
Here
$T_{\tau}$ is the time-ordering operator~\cite{AGD},
and $J_{S}^{x}$ and $J_{E}^{x}$ are
spin and energy, respectively, current operators. 
They are obtained from the continuity equations~\cite{Mahan,NA-ThCond1,NA-ThCond2}
(see Appendix A);
the results are
\begin{align}
  &J_{S}^{x}
  =-\sum_{\bdq}\sum_{l,l^{\prime}=A,B}v_{ll^{\prime}}^{x}(\bdq)x_{\bdq l}^{\dagger}x_{\bdq l^{\prime}},\label{eq:JS}\\
  &J_{E}^{x}
  =\sum_{\bdq}\sum_{l,l^{\prime}=A,B}e_{ll^{\prime}}^{x}(\bdq)x_{\bdq l}^{\dagger}x_{\bdq l^{\prime}},\label{eq:JE}
\end{align}
where $v_{ll^{\prime}}^{x}(\bdq)=(1-\delta_{l,l^{\prime}})
\frac{\partial \epsilon_{AB}(\bdq)}{\partial q_{x}}$,
$x_{\bdq A}=a_{\bdq}$, $x_{\bdq B}=b_{\bdq}^{\dagger}$,
$e_{BB}^{x}(\bdq)=-e_{AA}^{x}(\bdq)=\epsilon_{AB}(\bdq)
\frac{\partial \epsilon_{AB}(\bdq)}{\partial q_{x}}$, and 
$e_{AB}^{x}(\bdq)=e_{BA}^{x}(\bdq)=\frac{1}{2}(\epsilon_{AA}-\epsilon_{BB})
\frac{\partial \epsilon_{AB}(\bdq)}{\partial q_{x}}$.
In deriving Eqs. (\ref{eq:JS}) and (\ref{eq:JE}),
we have omitted the corrections due to $H_{\textrm{int}}$
because they may be negligible~\cite{Ogata}.
Then we can obtain $L_{11}$
by replacing $J_{E}^{x}$ in $\Phi_{12}(i\Omega_{n})$ by $J_{S}^{x}$,
and $L_{22}$ by replacing $J_{S}^{x}(\tau)$ in $\Phi_{12}(i\Omega_{n})$
by $J_{E}^{x}(\tau)$.
Thus the derivation of $L_{12}$ is enough
in obtaining $L_{\mu\nu}$'s.
In addition,
since we can derive $L_{12}$ 
in a similar way to the derivations of
electron transport coefficients~\cite{Eliashberg,NA-SCD,Kontani,Ogata,NA-ChTrans},
we explain its main points below.
(Note that the Bose-Einstein condensation of magnons is absent in our situation.)

By substituting Eqs. (\ref{eq:JS}) and (\ref{eq:JE}) into Eq. (\ref{eq:Phi12})
and performing some calculations (for the details see Appendix B),
we obtain
\begin{align}
  L_{12}=L_{12}^{0}+L_{12}^{\prime}.\label{eq:L12}
\end{align}
First, $L_{12}^{0}$, the noninteracting $L_{12}$, is given by (see Appendix B)
\begin{align}
  L_{12}^{0}
  &=\frac{1}{\pi N}\sum_{\bdq}\sum_{\nu,\nu^{\prime}=\alpha,\beta}
  v_{\nu^{\prime}\nu}^{x}(\bdq)e_{\nu\nu^{\prime}}^{x}(\bdq)
  I_{\nu\nu^{\prime}}^{(\textrm{I})}(\bdq),\label{eq:L12^0}
\end{align}
where
$v_{\nu^{\prime}\nu}^{x}(\bdq)
=\sum_{l,l^{\prime}=A,B}v_{ll^{\prime}}^{x}(\bdq)(U_{\bdq})_{l\nu^{\prime}}(U_{\bdq})_{l^{\prime}\nu}$,
$e_{\nu\nu^{\prime}}^{x}(\bdq)
=\sum_{l,l^{\prime}=A,B}e_{ll^{\prime}}^{x}(\bdq)(U_{\bdq})_{l\nu}(U_{\bdq})_{l^{\prime}\nu^{\prime}}$,
and
\begin{align}
  I_{\nu\nu^{\prime}}^{(\textrm{I})}(\bdq)
  =\int_{-\infty}^{\infty}dz\frac{\partial n(z)}{\partial z}
  \textrm{Im}G_{\nu}^{\textrm{R}}(\bdq,z)\textrm{Im}G_{\nu^{\prime}}^{\textrm{R}}(\bdq,z).\label{eq:I^I}
\end{align}
Here
$n(z)=(e^{z/T}-1)^{-1}$,
$G_{\alpha}^{\textrm{R}}(\bdq,z)=[z-\epsilon_{\alpha}(\bdq)+i\gamma]^{-1}$,
$G_{\beta}^{\textrm{R}}(\bdq,z)=-[z+\epsilon_{\beta}(\bdq)+i\gamma]^{-1}$,
and $\gamma$ is the magnon damping.
Next, 
$L_{12}^{\prime}$, the leading correction
due to the first-order perturbation of $H_{\textrm{int}}$, is given by (see Appendix B)
\begin{align}
  L_{12}^{\prime}
  =&\frac{1}{\pi^{2}N^{2}}\sum_{\bdq,\bdq^{\prime}}\sum_{\nu_{1}\nu_{2},\nu_{3},\nu_{4}}
  v_{\nu_{1}\nu_{2}}^{x}(\bdq)e_{\nu_{3}\nu_{4}}^{x}(\bdq^{\prime})
  V_{\nu_{1}\nu_{2}\nu_{3}\nu_{4}}(\bdq,\bdq^{\prime})\notag\\
  &\times 
  [I_{\nu_{1}\nu_{2}}^{(\textrm{I})}(\bdq)I_{\nu_{3}\nu_{4}}^{(\textrm{II})}(\bdq^{\prime})
  +I_{\nu_{1}\nu_{2}}^{(\textrm{II})}(\bdq)I_{\nu_{3}\nu_{4}}^{(\textrm{I})}(\bdq^{\prime})],\label{eq:L12'}
\end{align}
where
\begin{align}
  &I_{\nu\nu^{\prime}}^{(\textrm{II})}(\bdq)
  =\int_{-\infty}^{\infty}dz n(z)
  \textrm{Im}[G_{\nu}^{\textrm{R}}(\bdq,z)G_{\nu^{\prime}}^{\textrm{R}}(\bdq,z)],\label{eq:I^II}
\end{align}
$V_{\nu_{1}\nu_{2}\nu_{3}\nu_{4}}(\bdq,\bdq^{\prime})
=4J_{\bdq-\bdq^{\prime}}
\sum_{l}(U_{\bdq})_{l\nu_{1}}(U_{\bdq})_{\bar{l}\nu_{2}}(U_{\bdq^{\prime}})_{\bar{l}\nu_{3}}(U_{\bdq^{\prime}})_{l\nu_{4}}$,
and $\bar{l}$ is $B$ or $A$ for $l=A$ or $B$, respectively.
Then we obtain
\begin{align}
  L_{11}=L_{11}^{0}+L_{11}^{\prime},\
  L_{22}=L_{22}^{0}+L_{22}^{\prime},
\end{align}
where 
$L_{11}^{0}$, $L_{11}^{\prime}$, $L_{22}^{0}$, and $L_{22}^{\prime}$ are obtained
by replacing $e_{\nu\nu^{\prime}}^{x}(\bdq)$ in Eq. (\ref{eq:L12^0}) by $-v_{\nu\nu^{\prime}}^{x}(\bdq)$,
$e_{\nu_{3}\nu_{4}}^{x}(\bdq^{\prime})$ in Eq. (\ref{eq:L12'}) by $-v_{\nu_{3}\nu_{4}}^{x}(\bdq^{\prime})$,
$v_{\nu^{\prime}\nu}^{x}(\bdq)$ in Eq. (\ref{eq:L12^0}) by $-e_{\nu^{\prime}\nu}^{x}(\bdq)$,
and $v_{\nu_{1}\nu_{2}}^{x}(\bdq)$ in Eq. (\ref{eq:L12'}) by $-e_{\nu_{1}\nu_{2}}^{x}(\bdq)$,
respectively.

Since we suppose that the magnon lifetime $\tau=(2\gamma)^{-1}$ is
long enough to regard magnons as quasiparticles, 
we rewrite Eqs. (\ref{eq:L12^0}) and (\ref{eq:L12'})
by taking the limit $\tau\rightarrow \infty$.
First, 
Eq. (\ref{eq:L12^0}) reduces to
\begin{align}
  L_{12}^{0}\sim L_{12\alpha}^{0}+L_{12\beta}^{0},\label{eq:L12^0-approx}
\end{align}
where 
\begin{align}
  L_{12\nu}^{0}\sim \frac{1}{N}\sum_{\bdq}
  v_{\nu\nu}^{x}(\bdq)e_{\nu\nu}^{x}(\bdq)\tau
  \frac{\partial n[\epsilon_{\nu}(\bdq)]}{\partial \epsilon_{\nu}(\bdq)}.\label{eq:L12^0-approx-band}
\end{align}
(The detailed derivation is described in Appendix C.)
This expression is consistent with that obtained in the Boltzmann theory
with the relaxation-time approximation~\cite{SSE-theory}.
Equation (\ref{eq:L12^0-approx}) shows that
$L_{12}^{0}\approx L_{12\alpha}^{0}$ at sufficiently low temperatures for $S_{A}>S_{B}$
owing to 
$\frac{\partial n[\epsilon_{\alpha}(\bdq)]}{\partial \epsilon_{\alpha}(\bdq)}\gg\frac{\partial n[\epsilon_{\beta}(\bdq)]}{\partial \epsilon_{\beta}(\bdq)}$.
Similarly, 
we obtain
\begin{align}
  L_{11}^{0}\sim L_{11\alpha}^{0}+L_{11\beta}^{0},\
  L_{22}^{0}\sim L_{22\alpha}^{0}+L_{22\beta}^{0},
\end{align}
where $L_{11\nu}^{0}$ and $L_{22\nu}^{0}$ are obtained
by replacing $e_{\nu\nu}^{x}(\bdq)$ in Eq. (\ref{eq:L12^0-approx-band}) by $-v_{\nu\nu}^{x}(\bdq)$
and
by replacing $v_{\nu\nu}^{x}(\bdq)$ by $-e_{\nu\nu}^{x}(\bdq)$, respectively. 
Then, as we show in Appendix C,
Eq. (\ref{eq:L12'}) reduces to
\begin{align}
  L_{12}^{\prime}\sim
  L_{12\textrm{-intra}}^{\prime}+L_{12\textrm{-inter}1}^{\prime}+L_{12\textrm{-inter}2}^{\prime},\label{eq:L12'-approx}
\end{align}
where $L_{12\textrm{-intra}}^{\prime}$ is the correction due to the intraband interactions, 
\begin{align}
  L_{12\textrm{-intra}}^{\prime}
  =&\ L_{12\textrm{-intra-}\alpha}^{\prime}+L_{12\textrm{-intra-}\beta}^{\prime}\label{eq:L12'-intra-1},\\
  L_{12\textrm{-intra-}\nu}^{\prime}
  =&-\frac{2}{N^{2}}\sum_{\bdq,\bdq^{\prime}}
  v_{\nu\nu}^{x}(\bdq)e_{\nu\nu}^{x}(\bdq^{\prime})\tau
  V_{\nu\nu\nu\nu}(\bdq,\bdq^{\prime})\notag\\
  &\times 
  \frac{\partial n[\epsilon_{\nu}(\bdq)]}{\partial \epsilon_{\nu}(\bdq)}
  \frac{\partial n[\epsilon_{\nu}(\bdq^{\prime})]}{\partial \epsilon_{\nu}(\bdq^{\prime})},
  \label{eq:L12'-intra-2}
\end{align}
and $L_{12\textrm{-inter}1}^{\prime}$ and $L_{12\textrm{-inter}2}^{\prime}$ are
the corrections due to the interband interactions,
\begin{align}
  L_{12\textrm{-inter}1}^{\prime}
  =&-\frac{2}{N^{2}}\sum_{\bdq,\bdq^{\prime}}
  v_{\alpha\alpha}^{x}(\bdq)e_{\beta\beta}^{x}(\bdq^{\prime})\tau
  V_{\alpha\alpha\beta\beta}(\bdq,\bdq^{\prime})\notag\\
  &\times 
  \frac{\partial n[\epsilon_{\alpha}(\bdq)]}{\partial \epsilon_{\alpha}(\bdq)}
  \frac{\partial n[\epsilon_{\beta}(\bdq^{\prime})]}{\partial \epsilon_{\beta}(\bdq^{\prime})}\notag\\
  &-\frac{2}{N^{2}}\sum_{\bdq,\bdq^{\prime}}
  v_{\beta\beta}^{x}(\bdq)e_{\alpha\alpha}^{x}(\bdq^{\prime})\tau
  V_{\beta\beta\alpha\alpha}(\bdq,\bdq^{\prime})\notag\\
  &\times 
  \frac{\partial n[\epsilon_{\beta}(\bdq)]}{\partial \epsilon_{\beta}(\bdq)}
  \frac{\partial n[\epsilon_{\alpha}(\bdq^{\prime})]}{\partial \epsilon_{\alpha}(\bdq^{\prime})},
  \label{eq:L12'-inter1}\\
  L_{12\textrm{-inter}2}^{\prime}
  =&\ L_{12\textrm{-inter2-}\alpha}^{\prime}+L_{12\textrm{-inter2-}\beta}^{\prime}\notag\\
  =&\ (L_{\textrm{E}\alpha}^{\prime}+L_{\textrm{S}\alpha}^{\prime})
  +(L_{\textrm{E}\beta}^{\prime}+L_{\textrm{S}\beta}^{\prime}),\label{eq:L12'-inter2}\\
  L_{\textrm{E}\nu}^{\prime}
  =&\frac{2}{N^{2}}\sum_{\bdq,\bdq^{\prime}}
  v_{\nu\nu}^{x}(\bdq)e_{\alpha\beta}^{x}(\bdq^{\prime})\tau
  V_{\nu\nu\alpha\beta}(\bdq,\bdq^{\prime})\notag\\
  &\times \frac{\partial n[\epsilon_{\nu}(\bdq)]}{\partial \epsilon_{\nu}(\bdq)}
  \frac{n[\epsilon_{\alpha}(\bdq^{\prime})]-n[-\epsilon_{\beta}(\bdq^{\prime})]}
       {\epsilon_{\alpha}(\bdq^{\prime})+\epsilon_{\beta}(\bdq^{\prime})},\label{eq:L12_ED}\\
  L_{\textrm{S}\nu}^{\prime}
  =&\frac{2}{N^{2}}\sum_{\bdq,\bdq^{\prime}}
  v_{\alpha\beta}^{x}(\bdq)e_{\nu\nu}^{x}(\bdq^{\prime})\tau
  V_{\alpha\beta\nu\nu}(\bdq,\bdq^{\prime})\notag\\
  &\times \frac{n[\epsilon_{\alpha}(\bdq)]-n[-\epsilon_{\beta}(\bdq)]}
  {\epsilon_{\alpha}(\bdq)+\epsilon_{\beta}(\bdq)}
  \frac{\partial n[\epsilon_{\nu}(\bdq^{\prime})]}{\partial \epsilon_{\nu}(\bdq^{\prime})}.
  \label{eq:L12_SD}
\end{align}
Here the $V_{\nu_{1}\nu_{2}\nu_{3}\nu_{4}}(\bdq,\bdq^{\prime})$'s are given by
\begin{align}
  V_{\nu\nu\nu\nu}(\bdq,\bdq^{\prime})
  =&\ V_{\alpha\alpha\beta\beta}(\bdq,\bdq^{\prime})
  =V_{\beta\beta\alpha\alpha}(\bdq,\bdq^{\prime})\notag\\
  =&\ 2J_{\bdq-\bdq^{\prime}}\sinh2\theta_{\bdq}\sinh2\theta_{\bdq^{\prime}},\\
  V_{\nu\nu\alpha\beta}(\bdq,\bdq^{\prime})
  =&\ V_{\alpha\beta\nu\nu}(\bdq^{\prime},\bdq)\notag\\
  =&-2J_{\bdq-\bdq^{\prime}}\sinh2\theta_{\bdq}\cosh2\theta_{\bdq^{\prime}}.
\end{align}
Equation (\ref{eq:L12'-inter1}) shows that
the interband components of the magnon-magnon interactions 
cause
the energy-current-drag correction and the spin-current-drag correction, 
which are, in the case for $S_{A}>S_{B}$,
the first and the second term, respectively, of Eq. (\ref{eq:L12'-inter1}).
Furthermore, Eqs. (\ref{eq:L12_ED}) and (\ref{eq:L12_SD}) show that  
other interband components
cause the energy-current-drag corrections $L_{\textrm{E}\nu}^{\prime}$'s
and the spin-current-drag corrections $L_{\textrm{S}\nu}^{\prime}$'s.
Since these interband components cause
the interband momentum transfer,
$L_{12\textrm{-inter}1}^{\prime}$ and $L_{12\textrm{-inter}2}^{\prime}$
are the corrections due to the interband magnon drag.
The similar corrections are obtained for $L_{11}^{\prime}$ and $L_{22}^{\prime}$:
\begin{align}
  L_{11}^{\prime}&\sim
  L_{11\textrm{-intra}}^{\prime}+L_{11\textrm{-inter}1}^{\prime}+L_{11\textrm{-inter}2}^{\prime},\label{eq:L11'-approx}\\
  L_{22}^{\prime}&\sim
  L_{22\textrm{-intra}}^{\prime}+L_{22\textrm{-inter}1}^{\prime}+L_{22\textrm{-inter}2}^{\prime},\label{eq:L22'-approx}
\end{align}
where $L_{11\textrm{-intra}}^{\prime}$ and $L_{22\textrm{-intra}}^{\prime}$
are the corrections due to the intraband interactions, 
\begin{align}
  L_{11\textrm{-intra}}^{\prime}
  =&\ L_{11\textrm{-intra-}\alpha}^{\prime}+L_{11\textrm{-intra-}\beta}^{\prime}\label{eq:L11'-intra-1},\\
  L_{11\textrm{-intra-}\nu}^{\prime}
  =&\frac{2}{N^{2}}\sum_{\bdq,\bdq^{\prime}}
  v_{\nu\nu}^{x}(\bdq)v_{\nu\nu}^{x}(\bdq^{\prime})\tau
  V_{\nu\nu\nu\nu}(\bdq,\bdq^{\prime})\notag\\
  &\times 
  \frac{\partial n[\epsilon_{\nu}(\bdq)]}{\partial \epsilon_{\nu}(\bdq)}
  \frac{\partial n[\epsilon_{\nu}(\bdq^{\prime})]}{\partial \epsilon_{\nu}(\bdq^{\prime})},
  \label{eq:L11'-intra-2}\\
  L_{22\textrm{-intra}}^{\prime}
  =&\ L_{22\textrm{-intra-}\alpha}^{\prime}+L_{22\textrm{-intra-}\beta}^{\prime}\label{eq:L22'-intra-1},\\
  L_{22\textrm{-intra-}\nu}^{\prime}
  =&\frac{2}{N^{2}}\sum_{\bdq,\bdq^{\prime}}
  e_{\nu\nu}^{x}(\bdq)e_{\nu\nu}^{x}(\bdq^{\prime})\tau
  V_{\nu\nu\nu\nu}(\bdq,\bdq^{\prime})\notag\\
  &\times 
  \frac{\partial n[\epsilon_{\nu}(\bdq)]}{\partial \epsilon_{\nu}(\bdq)}
  \frac{\partial n[\epsilon_{\nu}(\bdq^{\prime})]}{\partial \epsilon_{\nu}(\bdq^{\prime})},
  \label{eq:L22'-intra-2}
\end{align}
and $L_{11\textrm{-inter}1}^{\prime}$, $L_{11\textrm{-inter}2}^{\prime}$,
$L_{22\textrm{-inter}1}^{\prime}$, and $L_{22\textrm{-inter}2}^{\prime}$
are the corrections due to the interband interactions,
\begin{align}
  L_{11\textrm{-inter}1}^{\prime}
  =& \frac{4}{N^{2}}\sum_{\bdq,\bdq^{\prime}}
  v_{\alpha\alpha}^{x}(\bdq)v_{\beta\beta}^{x}(\bdq^{\prime})\tau
  V_{\alpha\alpha\beta\beta}(\bdq,\bdq^{\prime})\notag\\
  &\times 
  \frac{\partial n[\epsilon_{\alpha}(\bdq)]}{\partial \epsilon_{\alpha}(\bdq)}
  \frac{\partial n[\epsilon_{\beta}(\bdq^{\prime})]}{\partial \epsilon_{\beta}(\bdq^{\prime})},
  \label{eq:L11'-inter1}\\
  L_{11\textrm{-inter}2}^{\prime}
  =&L_{11\textrm{-inter2-}\alpha}^{\prime}+L_{11\textrm{-inter2-}\beta}^{\prime},
  \label{eq:L11'-inter2-1}\\  
  L_{11\textrm{-inter2-}\nu}^{\prime}
  =&-\frac{4}{N^{2}}\sum_{\bdq,\bdq^{\prime}}
  v_{\nu\nu}^{x}(\bdq)v_{\alpha\beta}^{x}(\bdq^{\prime})\tau
  V_{\nu\nu\alpha\beta}(\bdq,\bdq^{\prime})\notag\\
  &\times \frac{\partial n[\epsilon_{\nu}(\bdq)]}{\partial \epsilon_{\nu}(\bdq)}
  \frac{n[\epsilon_{\alpha}(\bdq^{\prime})]-n[-\epsilon_{\beta}(\bdq^{\prime})]}
       {\epsilon_{\alpha}(\bdq^{\prime})+\epsilon_{\beta}(\bdq^{\prime})},
  \label{eq:L11'-inter2-2}\\
  L_{22\textrm{-inter}1}^{\prime}
  =& \frac{4}{N^{2}}\sum_{\bdq,\bdq^{\prime}}
  e_{\alpha\alpha}^{x}(\bdq)e_{\beta\beta}^{x}(\bdq^{\prime})\tau
  V_{\alpha\alpha\beta\beta}(\bdq,\bdq^{\prime})\notag\\
  &\times 
  \frac{\partial n[\epsilon_{\alpha}(\bdq)]}{\partial \epsilon_{\alpha}(\bdq)}
  \frac{\partial n[\epsilon_{\beta}(\bdq^{\prime})]}{\partial \epsilon_{\beta}(\bdq^{\prime})},
  \label{eq:L22'-inter1}\\
  L_{22\textrm{-inter}2}^{\prime}
  =&L_{22\textrm{-inter2-}\alpha}^{\prime}+L_{22\textrm{-inter2-}\beta}^{\prime},
  \label{eq:L22'-inter2-1}\\  
  L_{22\textrm{-inter2-}\nu}^{\prime}
  =&-\frac{4}{N^{2}}\sum_{\bdq,\bdq^{\prime}}
  e_{\nu\nu}^{x}(\bdq)e_{\alpha\beta}^{x}(\bdq^{\prime})\tau
  V_{\nu\nu\alpha\beta}(\bdq,\bdq^{\prime})\notag\\
  &\times \frac{\partial n[\epsilon_{\nu}(\bdq)]}{\partial \epsilon_{\nu}(\bdq)}
  \frac{n[\epsilon_{\alpha}(\bdq^{\prime})]-n[-\epsilon_{\beta}(\bdq^{\prime})]}
       {\epsilon_{\alpha}(\bdq^{\prime})+\epsilon_{\beta}(\bdq^{\prime})}.
  \label{eq:L22'-inter2-2}
\end{align}
As well as $L_{12\textrm{-inter}1}^{\prime}$ and $L_{12\textrm{-inter}2}^{\prime}$,
$L_{11\textrm{-inter}1}^{\prime}$, $L_{11\textrm{-inter}2}^{\prime}$,
$L_{22\textrm{-inter}1}^{\prime}$, and $L_{22\textrm{-inter}2}^{\prime}$
are the interband magnon drag corrections.

\section{Numerical results}

\begin{figure*}
  \includegraphics[width=180mm]{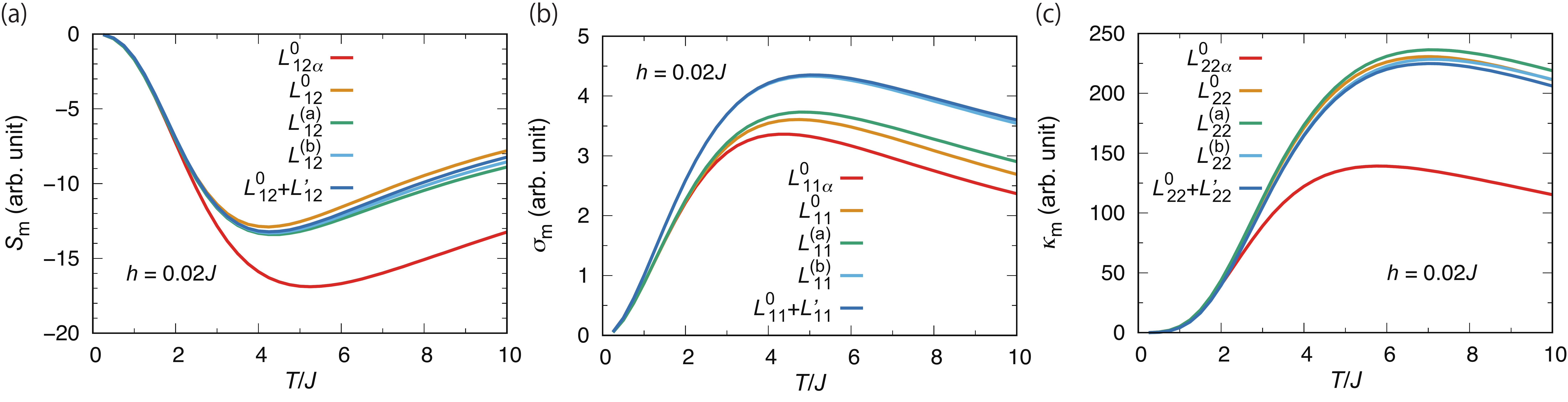}
  \caption{\label{fig2}
    The temperature dependences of (a) $S_{\textrm{m}}(=L_{12})$,
    (b) $\sigma_{\textrm{m}}(=L_{11})$, and
    (c) $\kappa_{\textrm{m}}(=L_{22})$ for $(S_{A},S_{B})=(\frac{3}{2},1)$ at $h=0.02J$.
    $L_{\mu\eta}^{(\textrm{a})}$ and $L_{\mu\eta}^{(\textrm{b})}$ 
    are defined as 
    $L_{\mu\eta}^{(\textrm{a})}=L_{\mu\eta}^{0}+L_{\mu\eta\textrm{-intra}}^{\prime}$
    and
    $L_{\mu\eta}^{(\textrm{b})}=L_{\mu\eta}^{0}+L_{\mu\eta\textrm{-intra}}^{\prime}+L_{\mu\eta\textrm{-inter}2}^{\prime}$,
    respectively.
    Note that $L_{\mu\eta}^{0}=L_{\mu\eta\alpha}^{0}+L_{\mu\eta\beta}^{0}$
    and
    $L_{\mu\eta}^{\prime}=L_{\mu\eta\textrm{-intra}}^{\prime}+L_{\mu\eta\textrm{-inter}1}^{\prime}+L_{\mu\eta\textrm{-inter}2}^{\prime}$.
    For $S_{\textrm{m}}$,
    the $L_{12\beta}^{0}$ is non-negligible for $T\geq 3J$
    and the largest term of the drag terms is $L_{12\textrm{-intra}}^{\prime}$,
    which enhances $|S_{\textrm{m}}|$.
    For $\sigma_{\textrm{m}}$, the $L_{11\beta}^{0}$ is non-negligible for $T\geq 4J$
    and the largest term of the drag terms is $L_{11\textrm{-inter2}}^{\prime}$,
    which enhances $\sigma_{\textrm{m}}$.
    For $\kappa_{\textrm{m}}$, the $L_{22\beta}^{0}$ is non-negligible for $T\geq 3J$
    and the largest term of the drag terms is $L_{22\textrm{-inter2}}^{\prime}$,
    which reduces $\kappa_{\textrm{m}}$.
    The effects of the other drag terms are summarized in Table \ref{tab}.
    }
\end{figure*}

\begin{table*}
  \caption{\label{tab}The effects of the drag terms
    on $L_{12}(=S_{\textrm{m}})$,
    $L_{11}(=\sigma_{\textrm{m}})$,
    and $L_{22}(=\kappa_{\textrm{m}})$.
    $|L_{12}|$ is enhanced by $L_{12\textrm{-intra}}^{\prime}$
    and reduced by $L_{12\textrm{-inter}1}^{\prime}$ and $L_{12\textrm{-inter}2}^{\prime}$.
    $L_{11}$ is enhanced by $L_{11\textrm{-intra}}^{\prime}$, 
    $L_{11\textrm{-inter}1}$, and $L_{11\textrm{-inter}2}^{\prime}$.
    $L_{22}$ is enhanced by $L_{22\textrm{-intra}}^{\prime}$
    and reduced by $L_{22\textrm{-inter}1}^{\prime}$ and $L_{22\textrm{-inter}2}^{\prime}$.
  }
  \begin{ruledtabular}
    \begin{tabular}{cccc}
      Transport coefficient & Intra term & Inter$1$ term & Inter$2$ term \\ \hline
      $|L_{12}|$ & Enhanced & Reduced & Reduced \\
      $L_{11}$ & Enhanced & Enhanced & Enhanced \\
      $L_{22}$ & Enhanced & Reduced & Reduced 
    \end{tabular}
  \end{ruledtabular}
\end{table*}

We numerically evaluate $S_{\textrm{m}}$,
$\sigma_{\textrm{m}}$, and $\kappa_{\textrm{m}}$.
We set $J=1$, $h=0.02J$, and $(S_{A},S_{B})=(\frac{3}{2},1)$.
$S_{A}:S_{B}=3:2$ is consistent with 
a ratio of Fe$^{\textrm{T}}$ to Fe$^{\textrm{O}}$ sites in the unit cell of YIG~\cite{YIG-1stPrinc}.
The reason why $(S_{A},S_{B})=(\frac{3}{2},1)$ is considered is that  
the transition temperature derived in a mean-field approximation in this case
with $J=3$ meV at $h=0$ [i.e., $T_{\textrm{c}}=(16/3)JS_{A}(S_{ B}+1)\sim 557$ K]
is close to the Curie temperature of YIG, $T_{\textrm{C}}$.
To perform the momentum summations numerically,
we divide the first Brillouin zone into a $N_{q}$-point mesh
and set $N_{q}=24^{3}(=N/2)$ (for more details, see Appendix D). 
The temperature range is chosen to be $0< T\leq 10J(\sim 0.6T_{\textrm{c}})$
because a previous study~\cite{Kanamori} showed that
the magnon theory in which the magnon-magnon interactions are considered 
in the first-order perturbation theory 
can reproduce the perpendicular spin susceptibility of MnF$_{2}$
up to about $0.6T_{\textrm{N}}$,
where $T_{\textrm{N}}$ is the N\'{e}el temperature.
For simplicity, we determine $\tau$ by $\tau^{-1}=\gamma_{0}+\gamma_{1}T+\gamma_{2}T^{2}$,
where $\gamma_{0}=10^{-2}J$, $\gamma_{1}=10^{-4}$, and $\gamma_{2}=10^{-3}$.
(The results shown below remain qualitatively unchanged at $h=0.08J$ and $0.16J$,
as shown in Appendix E.)

We begin with the temperature dependence of $S_{\textrm{m}}$.
Figure \ref{fig2}(a) shows that in the range of $0< T\leq 2J$ $L_{12}\approx L_{12\alpha}^{0}$ holds,
whereas for $T\geq 3J$ the contribution from $L_{12\beta}^{0}$ is non-negligible.
For example, at $T=6J$ we have $L_{12}^{0}/L_{12\alpha}^{0}\sim 0.7$.
This result indicates that
the higher-energy band magnons contribute to $S_{\textrm{m}}$ 
even for $T<[\epsilon_{\beta}(\bdq)-\epsilon_{\alpha}(\bdq)]=7.96J$.
This may be surprising
because their contributions are believed to be negligible
at such temperatures.
Then, Fig. \ref{fig2}(a) shows that
the magnitude of $S_{\textrm{m}}$ is enhanced by   
the intraband correction $L_{12\textrm{-intra}}^{\prime}[=L_{12}^{(a)}-L_{12}^{0}]$,
whereas
it is reduced by 
the interband corrections $L_{12\textrm{-inter}2}^{\prime}[=L_{12}^{(b)}-L_{12}^{(a)}]$
and $L_{12\textrm{-inter}1}^{\prime}[=L_{12}^{0}+L_{12}^{\prime}-L_{12}^{(b)}]$
(Table \ref{tab}).
Among these corrections,
$L_{12\textrm{-intra}}^{\prime}$ gives the largest contribution.
(As we will see below, this contrasts with the result of $L_{11}$ or $L_{22}$,
for which the largest contribution comes from $L_{11\textrm{-inter}2}^{\prime}$
or $L_{22\textrm{-inter}2}^{\prime}$, respectively.)
The reason why
the interband magnon drag corrections 
$L_{12\textrm{-inter}2}^{\prime}$ and $L_{12\textrm{-inter}1}^{\prime}$ are small
is that   
the energy-current-drag contributions and spin-current-drag contributions
[e.g., $L_{\textrm{E}\alpha}^{\prime}$ and $L_{\textrm{S}\alpha}^{\prime}$ in Eq. (\ref{eq:L12'-inter2})]
are opposite in sign and are nearly canceled out.
Figure \ref{fig2}(a) also shows 
$L_{12}^{0}+L_{12}^{\prime}\approx L_{12}^{0}$.
These results suggest that
the total effects of the interband magnon drag on $S_{\textrm{m}}$ are small.  

We turn to $\sigma_{\textrm{m}}$ and $\kappa_{\textrm{m}}$.
Their temperature dependences are shown in Figs. \ref{fig2}(b) and \ref{fig2}(c).
First, we see 
the $\beta$-band magnons contribute to $L_{11}$ for $T\geq 4J$ and to $L_{22}$ for $T\geq 3J$.
This result is similar to that of $L_{12}$ and indicates that
the multiband effects are significant also for $\sigma_{\textrm{m}}$ and $\kappa_{\textrm{m}}$.
The largest effects on $L_{22}$ are
due to
the property that $e_{\nu\nu}^{x}(\bdq)$ includes $\epsilon_{\nu}(\bdq)$
[more precisely, $e_{\alpha\alpha}^{x}(\bdq)=v_{\alpha\alpha}^{x}(\bdq)\epsilon_{\alpha}(\bdq)$
and $e_{\beta\beta}^{x}(\bdq)=-v_{\beta\beta}^{x}(\bdq)\epsilon_{\beta}(\bdq)$].
Then,
Figs. \ref{fig2}(b) and \ref{fig2}(c) show that
$\sigma_{\textrm{m}}$ is enhanced by 
$L_{11\textrm{-intra}}^{\prime}$,
$L_{11\textrm{-inter}2}^{\prime}$,
and $L_{11\textrm{-inter}1}^{\prime}$,
and that
$\kappa_{\textrm{m}}$ 
is enhanced by $L_{22\textrm{-intra}}^{\prime}$
and reduced by $L_{22\textrm{-inter}2}^{\prime}$
and $L_{22\textrm{-inter}1}^{\prime}$ (Table \ref{tab}).
[Note that $L_{\mu\eta\textrm{-intra}}^{\prime}=L_{\mu\eta}^{(a)}-L_{\mu\eta}^{0}$,
$L_{\mu\eta\textrm{-inter}2}^{\prime}=L_{\mu\eta}^{(b)}-L_{\mu\eta}^{(a)}$,
and $L_{\mu\eta\textrm{-inter}1}^{\prime}=L_{\mu\eta}^{0}+L_{\mu\eta}^{\prime}-L_{\mu\eta}^{(b)}$.]
In contrast to $L_{12}^{\prime}$,
the largest contributions to $L_{11}^{\prime}$ and $L_{22}^{\prime}$
come from $L_{11\textrm{-inter}2}^{\prime}$
and $L_{22\textrm{-inter}2}^{\prime}$, respectively.
Since $L_{11\textrm{-inter}2}^{\prime}$, $L_{11\textrm{-inter}1}^{\prime}$,
$L_{22\textrm{-inter}2}^{\prime}$, and $L_{22\textrm{-inter}1}^{\prime}$
are the interband magnon drag corrections,
the above results suggest that
the interband magnon drag enhances $\sigma_{\textrm{m}}$ and reduces $\kappa_{\textrm{m}}$.
This implies that
the interband magnon drag could be used
to enhance the spin current and 
to reduce the energy current.
Since this drag results from
the interband momentum transfer induced by the magnon-magnon interactions,
its effects could be controlled
by changing the band splitting energy
considerably via external fields.
(Such control is meaningful if and only if
the magnon picture remains valid.)
Note that for ferrimagnetic insulators
the effects of the weak magnetic field on the band splitting energy
are negligible because this energy for $h=0$
is of the order of $J$.
(The actual analysis about the possibility
of controlling the interband magnon drag
is a future problem.)

\section{Discussions}

We discuss the validity of our theory.
Since $H_{\textrm{int}}$ could be treated as perturbation except near $T_{\textrm{C}}$,
we believe 
our theory is appropriate for describing the magnon transport for $T<T_{\textrm{C}}$. 
It may be suitable to treat the magnon-magnon interactions in the Holstein-Primakoff method
because
the unphysical processes that can appear in a $S=1/2$ ferromagnet~\cite{Dyson}
are absent in our case.
Then
the effects of the magnon-phonon interactions may not change the results
qualitatively. 
First,
since the interaction-induced magnon polaron occurs
only at several values of $h$~\cite{MagPolaron}, 
its effect can be avoided.
Another effect 
is to cause the temperature dependence of $\tau$~\cite{MagPho-damp,Bauer-MagPhonDamp},
and 
it could be approximately considered as the temperature-dependent $\tau$.
Although the phonon-drag contributions might change $S_{\textrm{m}}$~\cite{PD-theory2},
experimental results~\cite{SSE-theory} suggest that
such contributions are small or negligible.

We make a short comment
about the relation between our theory and the Boltzmann theory.
Our theory is based on a method of Green's functions,
which can describe the effects of the damping and the vertex corrections
appropriately. 
In principle, these effects can be described also in the Boltzmann theory 
if the collision integral is treated appropriately~\cite{Boltzmann-CVC}.
However, in many analyses using the Boltzmann theory,
the collision integral is evaluated in the relaxation-time approximation,
in which the vertex corrections are completely omitted.
Since our interband magnon drag comes from the vertex corrections due to
the first-order perturbation of the quartic terms,
the similar result might be obtained also in the Boltzmann theory
if the interband components of the collision integral are treated appropriately.

We remark on the implications of our results.
First,
our interband magnon drag 
is distinct from a magnon drag in metals.
For the latter,
magnons drag an electron charge current 
via the second-order perturbation of a $sd$-type exchange interaction~\cite{MD-theory1}.
Second,
the interband magnon drag is possible in
various ferrimagnetic insulators
and other magnetic systems, such as 
antiferromagnets~\cite{AF-magnon,NA-AF,NA-ThCond1}
and spiral magnets~\cite{NA-ThCond2,NA-DM}.
Note that the possible ferrimagnetic insulators
include not only YIG,
but also some spinel ferrites, such as CoFe$_{2}$O$_{4}$
and NiFe$_{2}$O$_{4}$~\cite{ferrite1,ferrite2}.
Third,
our theory can be extended to phonons and photons.
Thus it may be useful for studying transport phenomena
of various interacting bosons.
Fourth,
our results will stimulate further studies of YIG.
For example,
the reduction in $|S_{\textrm{m}}|$ due to the multiband effect 
could improve the differences
between the voltages observed in the spin-Seebeck effect and
obtained in the Boltzmann theory of the ferromagnet~\cite{SSE-theory}
at high temperatures
because the voltage is proportional to $S_{\textrm{m}}$.

\section{Conclusion}

We have studied
$S_{\textrm{m}}$, $\sigma_{\textrm{m}}$, and $\kappa_{\textrm{m}}$
of interacting magnons in the minimal model of ferrimagnetic insulators.
We derived them by using the linear-response theory
and treating the magnon-magnon interactions as perturbation. 
We showed that
some interband components of the magnon-magnon interactions
give the corrections to these transport coefficients.
These corrections are due to the interband magnon drag,
which is distinct from the magnon drag in metals. 
Then we numerically calculated the temperature dependences of
$S_{\textrm{m}}$, $\sigma_{\textrm{m}}$, and $\kappa_{\textrm{m}}$
for $(S_{A},S_{B})=(\frac{3}{2},1)$ and $h=0.02J$.
We showed that
the total effects of the interband magnon drag on $S_{\textrm{m}}$ become small,
whereas it enhances $\sigma_{\textrm{m}}$ and reduces $\kappa_{\textrm{m}}$.
The latter result may suggest that 
the interband magnon drag could be used to enhance the spin current
and reduce the energy current.
For $S_{\textrm{m}}$,
the interband corrections become small
because
they lead to 
the energy-current-drag contributions and spin-current-drag contributions,
which are opposite in sign and are nearly canceled out.
We also showed that
the contributions from the higher-energy band magnons to
$S_{\textrm{m}}$, $\sigma_{\textrm{m}}$, and $\kappa_{\textrm{m}}$
are non-negligible 
even for temperatures lower than the band splitting.
This result indicates the importance of the multiband effects.

\begin{acknowledgments}
  This work was supported by JSPS KAKENHI Grants No. JP19K14664
  and JP22K03532.
  The author also acknowledges support from  
  JST CREST Grant No. JPMJCR1901.  
\end{acknowledgments}

\appendix

\section{Derivations of Eqs. (\ref{eq:JS}) and (\ref{eq:JE})}

We explain the details of the derivations
of $J_{S}^{x}$ and $J_{E}^{x}$, Eqs. (\ref{eq:JS}) and (\ref{eq:JE}).
As described in the main text,
they are obtained from
the continuity equations.
Such a derivation is explained, for example, in Ref. \onlinecite{Mahan}.

We begin with the derivation of $J_{S}^{x}$.
(Note that the following derivation,
which is applicable to collinear magnets, 
can be extended to
noncollinear magnets.)
We suppose that the $z$ component of a spin angular momentum, $S_{m}^{z}$,
satisfies 
\begin{align}
  \frac{d S_{m}^{z}}{d t}+\bdnabla\cdot\bdj_{m}^{(S)}=0,
\end{align}
where $\bdj_{m}^{(S)}$ is a spin current operator at site $m$.
Using this equation,
we have
\begin{align}
  \frac{d}{dt}\Bigl(\sum_{m}\bdR_{m}S_{m}^{z}\Bigr)
  &=-\sum_{m}\bdR_{m}\bdnabla\cdot\bdj_{m}^{(S)}\notag\\
  &=\sum_{m}\bdj_{m}^{(S)}
  =\bdJ^{(S)}_{l}.\label{eq:JS-start}
\end{align}
Here $l$ is $A$ or $B$
when the sum $\sum_{m}$ takes over
sites on the $A$ or the $B$ sublattice, respectively. 
In deriving the second equal in Eq. (\ref{eq:JS-start})
we have omitted the surface contributions.
$J_{S}^{x}$ is given by the $x$ component of $\bdJ_{S}$, where 
\begin{align}
  \bdJ_{S}=\bdJ_{A}^{(S)}+\bdJ_{B}^{(S)}.\label{eq:JS-total}
\end{align}
Combining Eq. (\ref{eq:JS-start}) with the Heisenberg equation of motion,
we obtain
\begin{align}
  \bdJ_{l}^{(S)}=i\sum_{m}\bdR_{m}\bigl[H,S_{m}^{z}\bigr],\label{eq:JS-Heis}
\end{align}
where $H$ is the Hamiltonian of the system considered.
Then,
since we focus on the magnon system described by $H=H_{\textrm{KE}}+H_{\textrm{int}}$,
where $H_{\textrm{KE}}$ and $H_{\textrm{int}}$ are given in the main text,
and treat $H_{\textrm{int}}$ as perturbation,
we replace $H$ in Eq. (\ref{eq:JS-Heis}) by $H_{\textrm{KE}}$
and $S_{m}^{z}$ in Eq. (\ref{eq:JS-Heis}) either by
$S_{A}-a_{m}^{\dagger}a_{m}$ for $l=A$
or by $-S_{B}+b_{m}^{\dagger}b_{m}$ for $l=B$;
as a result,
we obtain 
\begin{align}
  \bdJ_{A}^{(S)}
  &=i\sum_{\langle i,j\rangle}\sum_{m}\bdR_{m}
  \bigl[h_{ij}^{0},S_{A}-a_{m}^{\dagger}a_{m}\bigr],\label{eq:JS^A-start}\\
  \bdJ_{B}^{(S)}
  &=i\sum_{\langle i,j\rangle}\sum_{m}\bdR_{m}
  \bigl[h_{ij}^{0},-S_{B}+b_{m}^{\dagger}b_{m}\bigr],\label{eq:JS^B-start}
\end{align}
where $H_{\textrm{KE}}=\sum_{\langle i,j\rangle}h_{ij}^{0}$ and 
$h_{ij}^{0}=(2JS_{B}+\delta_{i,j}h)a_{i}^{\dagger}a_{i}+(2JS_{A}-\delta_{i,j}h)b_{j}^{\dagger}b_{j}+2J\sqrt{S_{A}S_{B}}(a_{i}^{\dagger}b_{j}^{\dagger}+a_{i}b_{j})$.
Note that the replacement of $H$ by $H_{\textrm{KE}}$ may be suitable
because the corrections due to $H_{\textrm{int}}$ are next-leading terms;
and that
the replacement of $S_{m}^{z}$ by $S_{A}-a_{m}^{\dagger}a_{m}$ 
or by $-S_{B}+b_{m}^{\dagger}b_{m}$ corresponds to
the Holstein-Primakoff transformation of the ferrimagnet.
After some algebra,
we can write Eqs. (\ref{eq:JS^A-start}) and (\ref{eq:JS^B-start}) as follows:
\begin{align}
  \bdJ_{A}^{(S)}
  &=-i2J\sqrt{S_{A}S_{B}}\sum_{\langle i,j\rangle}
  \bdR_{i}(a_{i}b_{j}-a_{i}^{\dagger}b_{j}^{\dagger}),\label{eq:JS^A-next}\\
  \bdJ_{B}^{(S)}
  &=i2J\sqrt{S_{A}S_{B}}\sum_{\langle i,j\rangle}
  \bdR_{j}(a_{i}b_{j}-a_{i}^{\dagger}b_{j}^{\dagger}).\label{eq:JS^B-next}
\end{align}
Combining these equations with Eq. (\ref{eq:JS-total}),
we have
\begin{align}
  \bdJ_{S}
  =-i2J\sqrt{S_{A}S_{B}}\sum_{\langle i,j\rangle}
  (\bdR_{i}-\bdR_{j})(a_{i}b_{j}-a_{i}^{\dagger}b_{j}^{\dagger}).\label{eq:JS-total-next}
\end{align}
Then, by using the Fourier coefficients of the magnon operators,
\begin{align}
  a_{i}=\sqrt{\frac{2}{N}}\sum_{\bdq}a_{\bdq}e^{i\bdq\cdot\bdR_{i}},\
  b_{j}^{\dagger}=\sqrt{\frac{2}{N}}\sum_{\bdq}b_{\bdq}^{\dagger}e^{i\bdq\cdot\bdR_{j}},\label{eq:Fourier}
\end{align}
we can rewrite Eq. (\ref{eq:JS-total-next}) as follows:
\begin{align}
  \bdJ_{S}
  &=-2J\sqrt{S_{A}S_{B}}\sum_{\bdq}
  \frac{\partial J_{\bdq}}{\partial \bdq}
  (a_{\bdq}b_{\bdq}+a_{\bdq}^{\dagger}b_{\bdq}^{\dagger})\notag\\
  &=-\sum_{\bdq}\frac{\partial \epsilon_{AB}(\bdq)}{\partial \bdq}
  (x_{\bdq B}^{\dagger}x_{\bdq A}+x_{\bdq A}^{\dagger}x_{\bdq B}),\label{eq:JS-last}
\end{align}
where
$J_{\bdq}=J\sum_{j=1}^{z}e^{i\bdq\cdot(\bdR_{i}-\bdR_{j})}=8J\cos\frac{q_{x}}{2}\cos\frac{q_{y}}{2}\cos\frac{q_{z}}{2}$,
$\epsilon_{AB}(\bdq)=2J\sqrt{S_{A}S_{B}}J_{\bdq}$,
$x_{\bdq A}=a_{\bdq}$, and $x_{\bdq B}=b_{\bdq}^{\dagger}$.
Note that $z$ is the number of nearest-neighbor sites ($z=8$). 
The $x$ component of Eq. (\ref{eq:JS-last}) gives Eq. (\ref{eq:JS}).

In a similar way we can obtain the expression of $J_{E}^{x}$.
(The following derivation is similar to
that for an antiferromagnet~\cite{NA-ThCond1}.)
First, we suppose that
the Hamiltonian at site $m$, $h_{m}$, satisfies 
\begin{align}
  \frac{d h_{m}}{dt}+\bdnabla\cdot\bdj_{m}^{(E)}=0,
\end{align}
where
$\bdj_{m}^{(E)}$ is an energy current operator at site $m$.
Because of this relation,
the energy current operator $\bdJ_{E}$ can be determined from
\begin{align}
  \bdJ_{E}=\bdJ^{(E)}_{A}+\bdJ^{(E)}_{B},\label{eq:JE-total}
\end{align}
where $\bdJ^{(E)}_{l}$ is given by
\begin{align}
  \bdJ^{(E)}_{l}
  =i\sum_{m,n}\bdR_{n}\bigl[h_{m},h_{n}\bigr],\label{eq:JE-Heis}
\end{align}
the sum $\sum_{m}$ take over sites on the $A$ or the $B$ sublattice,
and the sum $\sum_{n}$ take over sites on sublattice $l$.
Then, to calculate the commutator in Eq. (\ref{eq:JE-Heis}),
we consider the contributions only from $H_{\textrm{KE}}$ 
and neglect the corrections due to $H_{\textrm{int}}$, 
as in the derivation of $\bdJ^{(S)}_{l}$.
As a result,
$h_{m}$ for $m\in A$ is given by
\begin{align}
  &h_{mA}^{0}=(2S_{B}zJ+h)a_{m}^{\dagger}a_{m}
  +\sqrt{S_{A}S_{B}}\sum_{j}J_{mj}(a_{m}b_{j}+a_{m}^{\dagger}b_{j}^{\dagger}),\label{eq:h_mA}
\end{align}
and that for $m\in B$ is given by 
\begin{align}
  &h_{mB}^{0}=(2S_{A}zJ-h)b_{m}^{\dagger}b_{m}
  +\sqrt{S_{A}S_{B}}\sum_{i}J_{im}(a_{i}b_{m}+a_{i}^{\dagger}b_{m}^{\dagger}).\label{eq:h_mB}
\end{align}
Here $m\in A$ or $B$ means that
$m$ is on the $A$ or $B$ sublattice, respectively, 
and 
$J_{ij}=J_{ji}=J$ for nearest-neighbor sites $i$ and $j$.
Note that $\sum_{i=1}^{N/2}h_{iA}^{0}+\sum_{j=1}^{N/2}h_{jB}^{0}=H_{\textrm{KE}}$.
In our definition,
the energy current operator includes
the conribution from the Zeeman energy
[see Eq. (\ref{eq:JE-Heis}){--}(\ref{eq:h_mB})].
Combining Eqs. (\ref{eq:h_mA}) and (\ref{eq:h_mB})
with Eqs. (\ref{eq:JE-total}) and (\ref{eq:JE-Heis}),
we have
\begin{align}
  &\bdJ_{E}
  =i\sum_{m,n}\bdR_{n}\bigl[h_{mA}^{0},h_{nA}^{0}\bigr]
  +i\sum_{m,n}\bdR_{n}\bigl[h_{mB}^{0},h_{nB}^{0}\bigr]\notag\\
  &+i\sum_{m,n}\bdR_{n}\bigl[h_{mA}^{0},h_{nB}^{0}\bigr]
  +i\sum_{m,n}\bdR_{n}\bigl[h_{mB}^{0},h_{nA}^{0}\bigr].\label{eq:JE-start}
\end{align}
Then we can calculate the commutators in Eq. (\ref{eq:JE-start}) by using
the commutation relations of the magnon operators and 
the identities $[AB,C]=A[B,C]+[A,C]B$ and $[A,BC]=[A,B]C+B[A,C]$;
the results are
\begin{align}
  \bigl[h_{mA}^{0},h_{nA}^{0}\bigr]
  &=S_{A}S_{B}\sum_{j}J_{mj}J_{nj}(a_{n}^{\dagger}a_{m}-a_{m}^{\dagger}a_{n}),\label{eq:h_A-h_A}\\
  \bigl[h_{mB}^{0},h_{nB}^{0}\bigr]
  &=S_{A}S_{B}\sum_{i}J_{im}J_{in}(b_{m}b_{n}^{\dagger}-b_{n}b_{m}^{\dagger}),\label{eq:h_B-h_B}\\
  \bigl[h_{mA}^{0},h_{nB}^{0}\bigr]
  &=S_{A}S_{B}\sum_{j}J_{mn}J_{mj}(b_{j}b_{n}^{\dagger}-b_{n}b_{j}^{\dagger})\notag\\
  &+[2Jz(S_{A}-S_{B})-2h]\sqrt{S_{A}S_{B}}J_{mn}\notag\\
  &\times (a_{m}b_{n}-a_{m}^{\dagger}b_{n}^{\dagger})\notag\\
  &+S_{A}S_{B}\sum_{i}J_{mn}J_{ni}(a_{i}^{\dagger}a_{m}-a_{m}^{\dagger}a_{i}),\label{eq:h_A-h_B}\\
  \bigl[h_{mB}^{0},h_{nA}^{0}\bigr]
  &=S_{A}S_{B}\sum_{j}J_{mn}J_{nj}(b_{m}b_{j}^{\dagger}-b_{j}b_{m}^{\dagger})\notag\\
  &+[-2Jz(S_{A}-S_{B})+2h]\sqrt{S_{A}S_{B}}J_{nm}\notag\\
  &\times (a_{n}b_{m}-a_{n}^{\dagger}b_{m}^{\dagger})\notag\\
  &+S_{A}S_{B}\sum_{i}J_{nm}J_{mi}(a_{n}^{\dagger}a_{i}-a_{i}^{\dagger}a_{n}).\label{eq:h_B-h_A}
\end{align}
By substituting these equations into Eq. (\ref{eq:JE-start})
and performing some calculations,
we obtain
\begin{align}
  \bdJ_{E}
  &=2i\sum_{m,n,j}(\bdR_{n}-\bdR_{m})S_{A}S_{B}J_{nj}J_{jm}a_{n}^{\dagger}a_{m}\notag\\
  &-2i\sum_{m,n,i}(\bdR_{n}-\bdR_{m})S_{A}S_{B}J_{ni}J_{im}b_{n}b_{m}^{\dagger}\notag\\
  &+i\sum_{m,n}(\bdR_{n}-\bdR_{m})[2Jz(S_{A}-S_{B})-2h]\sqrt{S_{A}S_{B}}\notag\\
  &\times 
  J_{mn}(a_{m}b_{n}-a_{m}^{\dagger}b_{n}^{\dagger}).\label{eq:JE-next}
\end{align}
As in the derivation of $\bdJ_{S}$,
we can rewrite Eq. (\ref{eq:JE-next}) by using
the Fourier coefficients of the magnon operators [Eq. (\ref{eq:Fourier})];
as a result, we have
\begin{align}
  \bdJ_{E}
  &=-\sum_{\bdq}2\sqrt{S_{A}S_{B}}J_{\bdq}2\sqrt{S_{A}S_{B}}\frac{\partial J_{\bdq}}{\partial \bdq}
  (a_{\bdq}^{\dagger}a_{\bdq}-b_{\bdq}b_{\bdq}^{\dagger})\notag\\
  &-[J_{\bdzero}(S_{A}-S_{B})-h]2\sqrt{S_{A}S_{B}}\frac{\partial J_{\bdq}}{\partial \bdq}
  (a_{\bdq}b_{\bdq}+a_{\bdq}^{\dagger}b_{\bdq}^{\dagger}).\label{eq:JE-nextnext}
\end{align}
Since
$\epsilon_{AA}=2J_{\bdzero}S_{B}+h$, $\epsilon_{BB}=2J_{\bdzero}S_{A}-h$,
and 
$\epsilon_{AB}(\bdq)=2\sqrt{S_{A}S_{B}}J_{\bdq}$,
we can write Eq. (\ref{eq:JE-nextnext}) as follows:
\begin{align}
  \bdJ_{E}
  &=-\sum_{\bdq}\epsilon_{AB}(\bdq)\frac{\partial \epsilon_{AB}(\bdq)}{\partial \bdq}
  (a_{\bdq}^{\dagger}a_{\bdq}-b_{\bdq}b_{\bdq}^{\dagger})\notag\\
  &+\sum_{\bdq}\frac{1}{2}(\epsilon_{AA}-\epsilon_{BB})
  \frac{\partial \epsilon_{AB}(\bdq)}{\partial \bdq}
  (a_{\bdq}b_{\bdq}+a_{\bdq}^{\dagger}b_{\bdq}^{\dagger})\notag\\
  &=\sum_{\bdq}\sum_{l,l^{\prime}=A,B}\bde_{ll^{\prime}}(\bdq)x_{\bdq l}^{\dagger}x_{\bdq l^{\prime}},\label{eq:JE-last}
\end{align}
where $\bde_{AA}(\bdq)=-\bde_{BB}(\bdq)=-\epsilon_{AB}(\bdq)\frac{\partial \epsilon_{AB}(\bdq)}{\partial \bdq}$ and 
$\bde_{AB}(\bdq)=\bde_{BA}(\bdq)=\frac{1}{2}(\epsilon_{AA}-\epsilon_{BB})\frac{\partial \epsilon_{AB}(\bdq)}{\partial \bdq}$.
Equation (\ref{eq:JE-last}) for the $x$ component is Eq. (\ref{eq:JE}).

\section{Derivations of Eqs. (\ref{eq:L12^0}) and (\ref{eq:L12'})}

We derive Eqs. (\ref{eq:L12^0}) and (\ref{eq:L12'}).
As described in the main text,
their derivations can be done in a similar way to the derivations of
electron transport coefficients~\cite{Eliashberg,Kontani,NA-ChTrans,Ogata}:
the transport coefficients can be derived by using
a method of Green's functions~\cite{AGD}. 
We first derive $L_{12}^{0}$, the noninteracting $L_{12}$, 
and then derive $L_{12}^{\prime}$,
the leading correction to $L_{12}$ due to the first-order perturbation of $H_{\textrm{int}}$.

\begin{figure*}
  \includegraphics[width=180mm]{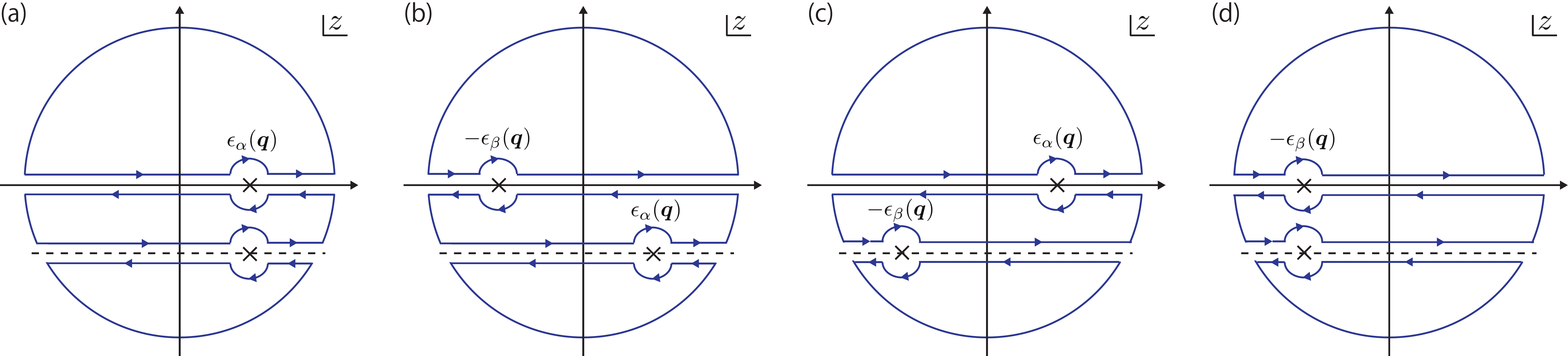}
  \caption{\label{figS1}
    The contours used for the integrations in
    (a) $G_{\alpha\alpha}^{(\textrm{II})}(\bdq,i\Omega_{n})$,
    (b) $G_{\alpha\beta}^{(\textrm{II})}(\bdq,i\Omega_{n})$,
    (c) $G_{\beta\alpha}^{(\textrm{II})}(\bdq,i\Omega_{n})$, 
    and (d) $G_{\beta\beta}^{(\textrm{II})}(\bdq,i\Omega_{n})$.
    The horizontal dashed lines correspond to $\textrm{Im}z=-\Omega_{n}$.
  }
\end{figure*}

First, we derive $L_{12}^{0}$, Eq. (\ref{eq:L12^0}).
Substituting Eqs. (\ref{eq:JS}) and (\ref{eq:JE}) into Eq. (\ref{eq:Phi12}),
we have
\begin{align}
  \Phi_{12}(i\Omega_{n})
  &=-\frac{1}{N}\sum_{\bdq,\bdq^{\prime}}\sum_{l_{1},l_{2},l_{3},l_{4}=A,B}
  v_{l_{1}l_{2}}^{x}(\bdq)e_{l_{3}l_{4}}^{x}(\bdq^{\prime})\notag\\
  &\times \int_{0}^{T^{-1}}d\tau e^{i\Omega_{n}\tau}
  \langle T_{\tau}x_{\bdq l_{1}}^{\dagger}(\tau)x_{\bdq l_{2}}(\tau)
  x_{\bdq^{\prime}l_{3}}^{\dagger}x_{\bdq^{\prime}l_{4}}\rangle\notag\\
  &=-\frac{1}{N}\sum_{\bdq,\bdq^{\prime}}\sum_{l_{1},l_{2},l_{3},l_{4}=A,B}
  v_{l_{1}l_{2}}^{x}(\bdq)e_{l_{3}l_{4}}^{x}(\bdq^{\prime})\notag\\
  &\times
  G^{(\textrm{II})}_{l_{1}l_{2}l_{3}l_{4}}(\bdq,\bdq^{\prime};i\Omega_{n}),\label{eq:Phi12-start}
\end{align}
where $\Omega_{n}=2\pi T n$ with $n>0$.
(Note that
the $n$ and $m$ used in this section are different from those used in Appendix A.) 
Equation (\ref{eq:Phi12-start}) provides a starting point to
derive $L_{12}^{0}$ and $L_{12}^{\prime}$.
To derive $L_{12}^{0}$,
we calculate $G^{(\textrm{II})}_{l_{1}l_{2}l_{3}l_{4}}(\bdq,\bdq^{\prime};i\Omega_{n})$
in the absence of $H_{\textrm{int}}$ by using Wick's theorem~\cite{AGD};
the result is
\begin{align}
  G^{(\textrm{II})}_{l_{1}l_{2}l_{3}l_{4}}(\bdq,\bdq^{\prime};i\Omega_{n})
  =&\delta_{\bdq,\bdq^{\prime}}T\sum_{m}
  G_{l_{2}l_{3}}(\bdq,i\Omega_{n}+i\Omega_{m})\notag\\
  &\times G_{l_{4}l_{1}}(\bdq,i\Omega_{m}),\label{eq:G^II0}
\end{align}
where $G_{ll^{\prime}}(\bdq,i\Omega_{m})$ is the magnon Green's function
in the sublattice basis 
with $\Omega_{m}=2\pi T m$ and an integer $m$,
\begin{align}
  G_{ll^{\prime}}(\bdq,i\Omega_{m})
  =-\int_{0}^{T^{-1}}d\tau e^{i\Omega_{m}\tau}
  \langle T_{\tau}x_{\bdq l}(\tau)x_{\bdq l^{\prime}}^{\dagger}\rangle.
\end{align}
Then the magnon operators in the sublattice basis, $x_{\bdq l}$ and $x_{\bdq l}^{\dagger}$,
are connected with those in the band basis, $x_{\bdq \nu}$ and $x_{\bdq \nu}^{\dagger}$, 
through the Bogoliubov transformation,
\begin{align}
  x_{\bdq l}=\sum_{\nu=\alpha,\beta}(U_{\bdq})_{l\nu}x_{\bdq\nu},\label{eq:Bogo}
\end{align}
where $x_{\bdq\alpha}=\alpha_{\bdq}$, $x_{\bdq\beta}=\beta_{\bdq}^{\dagger}$, 
$(U_{\bdq})_{A\alpha}=(U_{\bdq})_{B\beta}=\cosh\theta_{\bdq}$, and 
$(U_{\bdq})_{A\beta}=(U_{\bdq})_{B\alpha}=-\sinh\theta_{\bdq}$;
as described in the main text, 
these hyperbolic functions satisfy
$\cosh2\theta_{\bdq}=\frac{J_{\bdzero}(S_{A}+S_{B})}{\Delta\epsilon_{\bdq}}$
and
$\sinh2\theta_{\bdq}=\frac{2\sqrt{S_{A}S_{B}}J_{\bdq}}{\Delta\epsilon_{\bdq}}$. 
Thus $G_{ll^{\prime}}(\bdq,i\Omega_{m})$ is related to the magnon Green's function in the band basis,
$G_{\nu}(\bdq,i\Omega_{m})$: 
\begin{align}
  G_{ll^{\prime}}(\bdq,i\Omega_{m})
  =\sum_{\nu=\alpha,\beta}(U_{\bdq})_{l\nu}(U_{\bdq})_{l^{\prime}\nu}
  G_{\nu}(\bdq,i\Omega_{m}),\label{eq:G_sub-band}
\end{align}
where
\begin{align}
  G_{\alpha}(\bdq,i\Omega_{m})=\frac{1}{i\Omega_{m}-\epsilon_{\alpha}(\bdq)},\
  G_{\beta}(\bdq,i\Omega_{m})=-\frac{1}{i\Omega_{m}+\epsilon_{\beta}(\bdq)}.\label{eq:G-band}
\end{align}
Combining Eq. (\ref{eq:G_sub-band}) with Eqs. (\ref{eq:G^II0}) and (\ref{eq:Phi12-start}),
we have
\begin{align}
  \Phi_{12}(i\Omega_{n})
  &=-\frac{1}{N}\sum_{\bdq}\sum_{\nu,\nu^{\prime}=\alpha,\beta}
  v_{\nu^{\prime}\nu}^{x}(\bdq)e_{\nu\nu^{\prime}}^{x}(\bdq)\notag\\
  &\times T\sum_{m}G_{\nu}(\bdq,i\Omega_{n+m})G_{\nu^{\prime}}(\bdq,i\Omega_{m})\notag\\
  &=-\frac{1}{N}\sum_{\bdq}\sum_{\nu,\nu^{\prime}=\alpha,\beta}
  v_{\nu^{\prime}\nu}^{x}(\bdq)e_{\nu\nu^{\prime}}^{x}(\bdq)
  G_{\nu\nu^{\prime}}^{(\textrm{II})}(\bdq,i\Omega_{n}),\label{eq:Phi12-next}
\end{align}
where
\begin{align}
  v_{\nu^{\prime}\nu}^{x}(\bdq)
  &=\sum_{l_{1},l_{2}=A,B}v_{l_{1}l_{2}}^{x}(\bdq)(U_{\bdq})_{l_{1}\nu^{\prime}}(U_{\bdq})_{l_{2}\nu},\\
  e_{\nu\nu^{\prime}}^{x}(\bdq)
  &=\sum_{l_{3},l_{4}=A,B}e_{l_{3}l_{4}}^{x}(\bdq)(U_{\bdq})_{l_{3}\nu}(U_{\bdq})_{l_{4}\nu^{\prime}}.
\end{align}
Then
we can rewrite $G_{\nu\nu^{\prime}}^{(\textrm{II})}(\bdq,i\Omega_{n})$ in Eq. (\ref{eq:Phi12-next}) 
as follows:
\begin{align}
  &G_{\nu\nu^{\prime}}^{(\textrm{II})}(\bdq,i\Omega_{n})
  =\int_{\textrm{C}}\frac{dz}{2\pi i}n(z)
  G_{\nu}(\bdq,i\Omega_{n}+z)G_{\nu^{\prime}}(\bdq,z)\notag\\
  &+T[G_{\nu}(\bdq,i\Omega_{n})G_{\nu^{\prime}}(\bdq,0)
  +G_{\nu}(\bdq,0)G_{\nu^{\prime}}(\bdq,-i\Omega_{n})],\label{eq:G^II0-next}
\end{align}
where $n(z)$ is the Bose distribution function, $n(z)=(e^{z/T}-1)^{-1}$,
and $\textrm{C}$ is one of the contours shown in Fig. \ref{figS1}.
Using Eqs. (\ref{eq:G^II0-next}) and (\ref{eq:G-band}),
we obtain
\begin{widetext}
\begin{align}
  G_{\nu\nu^{\prime}}^{(\textrm{II})}(\bdq,i\Omega_{n})
  =\int_{-\infty}^{\infty}\frac{dz}{2\pi i}n(z)
  \Bigl\{
  G_{\nu}^{\textrm{R}}(\bdq,z+i\Omega_{n})
  [G_{\nu^{\prime}}^{\textrm{R}}(\bdq,z)-G_{\nu^{\prime}}^{\textrm{A}}(\bdq,z)]
  +[G_{\nu}^{\textrm{R}}(\bdq,z)-G_{\nu}^{\textrm{A}}(\bdq,z)]
  G_{\nu^{\prime}}^{\textrm{A}}(\bdq,z-i\Omega_{n})
  \Bigr\},\label{eq:G^II0-next-next}
\end{align}
where $G_{\nu}^{\textrm{R}}(\bdq,z)$ is the retarded magnon Green's function,
\begin{align}
  G_{\alpha}^{\textrm{R}}(\bdq,z)=\frac{1}{z-\epsilon_{\alpha}(\bdq)+i\gamma},\
  G_{\beta}^{\textrm{R}}(\bdq,z)=-\frac{1}{z+\epsilon_{\beta}(\bdq)+i\gamma},\label{eq:G^R-band}
\end{align}
$G_{\nu}^{\textrm{A}}(\bdq,z)$ is the advanced one, 
and $\gamma$ is the magnon damping. 
By combining Eq. (\ref{eq:G^II0-next-next}) with Eq. (\ref{eq:Phi12-next}) and
performing the analytic continuation $i\Omega_{n}\rightarrow \omega+i\delta$ with $\delta=0+$, 
we have
\begin{align}
  \Phi_{12}^{\textrm{R}}(\omega)
  =\Phi_{12}(i\Omega_{n}\rightarrow \omega+i\delta)
  =&-\frac{1}{N}\sum_{\bdq}\sum_{\nu,\nu^{\prime}=\alpha,\beta}
  v_{\nu^{\prime}\nu}^{x}(\bdq)e_{\nu\nu^{\prime}}^{x}(\bdq)
  \int_{-\infty}^{\infty}\frac{dz}{2\pi i}n(z)\notag\\
  &\times
  \Bigl\{
  G_{\nu}^{\textrm{R}}(\bdq,z+\omega)
  [G_{\nu^{\prime}}^{\textrm{R}}(\bdq,z)-G_{\nu^{\prime}}^{\textrm{A}}(\bdq,z)]
  +[G_{\nu}^{\textrm{R}}(\bdq,z)-G_{\nu}^{\textrm{A}}(\bdq,z)]
  G_{\nu^{\prime}}^{\textrm{A}}(\bdq,z-\omega)
  \Bigr\}.
\end{align}
By using $G(z+\omega)=G(z)+\omega\frac{\partial G(z)}{\partial z}+O(\omega^{2})$
and performing the partial integration,
we obtain
\begin{align}
  L_{12}^{0}
  =&\lim_{\omega\rightarrow 0}\frac{\Phi_{12}^{\textrm{R}}(\omega)-\Phi_{12}^{\textrm{R}}(0)}{i\omega}\notag\\
  =&-\frac{1}{4N}\sum_{\bdq}\sum_{\nu,\nu^{\prime}=\alpha,\beta}
  v_{\nu^{\prime}\nu}^{x}(\bdq)e_{\nu\nu^{\prime}}^{x}(\bdq)
  \int_{-\infty}^{\infty}\frac{dz}{\pi}\frac{\partial n(z)}{\partial z}
  \Bigl[
    G_{\nu}^{\textrm{R}}(\bdq,z)G_{\nu^{\prime}}^{\textrm{R}}(\bdq,z)
    -2G_{\nu}^{\textrm{R}}(\bdq,z)G_{\nu^{\prime}}^{\textrm{A}}(\bdq,z)
    +G_{\nu}^{\textrm{A}}(\bdq,z)G_{\nu^{\prime}}^{\textrm{A}}(\bdq,z)
    \Bigr]\notag\\
  =&\frac{1}{N}\sum_{\bdq}\sum_{\nu,\nu^{\prime}=\alpha,\beta}
  v_{\nu^{\prime}\nu}^{x}(\bdq)e_{\nu\nu^{\prime}}^{x}(\bdq)
  \int_{-\infty}^{\infty}\frac{dz}{\pi}\frac{\partial n(z)}{\partial z}
  \textrm{Im}G_{\nu}^{\textrm{R}}(\bdq,z)\textrm{Im}G_{\nu^{\prime}}^{\textrm{R}}(\bdq,z).\label{eq:L12^0-A}
\end{align}
In deriving this equation
we have used the symmetry relations
$v_{\nu^{\prime}\nu}^{x}(\bdq)=v_{\nu\nu^{\prime}}^{x}(\bdq)$
and $e_{\nu\nu^{\prime}}^{x}(\bdq)=e_{\nu^{\prime}\nu}^{x}(\bdq)$. 
Equation (\ref{eq:L12^0-A}) is Eq. (\ref{eq:L12^0}).

Next, we derive $L_{12}^{\prime}$, Eq. (\ref{eq:L12'}).
By using Eq. (\ref{eq:Phi12-start}),
we can write the correction due to the first-order perturbation of $H_{\textrm{int}}$ as follows:
\begin{align}
  \hspace{-10pt}
  \Delta\Phi_{12}(i\Omega_{n})
  =+\frac{1}{N}\sum_{\bdq,\bdq^{\prime}}\sum_{l_{1},l_{2},l_{3},l_{4}=A,B}
  v_{l_{1}l_{2}}^{x}(\bdq)e_{l_{3}l_{4}}^{x}(\bdq^{\prime})
  \int_{0}^{T^{-1}}d\tau e^{i\Omega_{n}\tau}\int_{0}^{T^{-1}}d\tau_{1}
  \langle T_{\tau}x_{\bdq l_{1}}^{\dagger}(\tau)x_{\bdq l_{2}}(\tau)
  x_{\bdq^{\prime}l_{3}}^{\dagger}x_{\bdq^{\prime}l_{4}}H_{\textrm{int}}(\tau_{1})\rangle.\label{eq:Phi12-int-start}
\end{align}
[Note that 
$H_{\textrm{int}}$ has been defined in Eq. (\ref{eq:Hint}).]
By using Wick's theorem~\cite{AGD},
we can calculate
$\langle T_{\tau}x_{\bdq l_{1}}^{\dagger}(\tau)x_{\bdq l_{2}}(\tau)x_{\bdq^{\prime}l_{3}}^{\dagger}x_{\bdq^{\prime}l_{4}}H_{\textrm{int}}(\tau_{1})\rangle$;
the result is
\begin{align}
  \langle T_{\tau}x_{\bdq l_{1}}^{\dagger}(\tau)x_{\bdq l_{2}}(\tau)
  x_{\bdq^{\prime}l_{3}}^{\dagger}x_{\bdq^{\prime}l_{4}}H_{\textrm{int}}(\tau_{1})\rangle
  =&-\frac{1}{N}\sum_{l_{5},l_{6},l_{7},l_{8}=A,B}
  V_{l_{5}l_{6}l_{7}l_{8}}(\bdq,\bdq^{\prime})
  G_{l_{5}l_{1}}(\bdq,\tau_{1}-\tau)G_{l_{2}l_{6}}(\bdq,\tau-\tau_{1})\notag\\
  &\times G_{l_{7}l_{3}}(\bdq^{\prime},\tau_{1})G_{l_{4}l_{8}}(\bdq^{\prime},-\tau_{1}),\label{eq:G^II-int}
\end{align}
where $G_{ll^{\prime}}(\bdq,\tau)=T\sum_{m}e^{-i\Omega_{m}\tau}G_{ll^{\prime}}(\bdq,i\Omega_{m})$,
\begin{align}
  V_{l_{5}l_{6}l_{7}l_{8}}(\bdq,\bdq^{\prime})
  &=\begin{cases}
  4J_{\bdzero}\ \ \ \ \ \ \ \ \ \ (l_{5}=l_{6}=l,l_{7}=l_{8}=\bar{l}),\\
  4J_{\bdq-\bdq^{\prime}}\ \ \ \ \ \ (l_{5}=l_{8}=l,l_{6}=l_{7}=\bar{l}),\\
  2J_{\bdq^{\prime}}\sqrt{\frac{S_{A}}{S_{B}}}\ \ (l_{5}=l_{6}=B,l_{7}=l,l_{8}=\bar{l}),\\
  2J_{\bdq}\sqrt{\frac{S_{A}}{S_{B}}}\ \ \ (l_{5}=l,l_{6}=\bar{l},l_{7}=l_{8}=B),\\
  2J_{\bdq^{\prime}}\sqrt{\frac{S_{B}}{S_{A}}}\ \ (l_{5}=l_{6}=A,l_{7}=l,l_{8}=\bar{l}),\\
  2J_{\bdq}\sqrt{\frac{S_{B}}{S_{A}}}\ \ \ (l_{5}=l,l_{6}=\bar{l},l_{7}=l_{8}=A),
  \end{cases}\label{eq:Vq}
\end{align}
and $\bar{l}=B$ or $A$ for $l=A$ or $B$, respectively.
Then,
by substituting Eq. (\ref{eq:G^II-int}) into Eq. (\ref{eq:Phi12-int-start})
and carrying out the integrations,
we obtain
\begin{align}
  \Delta\Phi_{12}(i\Omega_{n})
  =&-\frac{1}{N^{2}}\sum_{\bdq,\bdq^{\prime}}\sum_{l_{1},l_{2},\cdots,l_{8}=A,B}
  v_{l_{1}l_{2}}^{x}(\bdq)e_{l_{3}l_{4}}^{x}(\bdq^{\prime})
  V_{l_{5}l_{6}l_{7}l_{8}}(\bdq,\bdq^{\prime})
  T^{2}\sum_{m,m^{\prime}}
  G_{l_{5}l_{1}}(\bdq,i\Omega_{m})G_{l_{2}l_{6}}(\bdq,i\Omega_{n+m})\notag\\
  &\times
  G_{l_{7}l_{3}}(\bdq^{\prime},i\Omega_{n+m^{\prime}})G_{l_{4}l_{8}}(\bdq^{\prime},i\Omega_{m^{\prime}}).
\end{align}
Furthermore,
we can rewrite this equation by using the Bogoliubov transformation [i.e., Eq. (\ref{eq:Bogo})];
the result is
\begin{align}
  \Delta\Phi_{12}(i\Omega_{n})
  =&-\frac{1}{N^{2}}\sum_{\bdq,\bdq^{\prime}}\sum_{\nu_{1},\nu_{2},\nu_{3},\nu_{4}=\alpha,\beta}
  v_{\nu_{1}\nu_{2}}^{x}(\bdq)e_{\nu_{3}\nu_{4}}^{x}(\bdq^{\prime})
  V_{\nu_{1}\nu_{2}\nu_{3}\nu_{4}}(\bdq,\bdq^{\prime})
  \Delta G_{\nu_{1}\nu_{2}\nu_{3}\nu_{4}}^{(\textrm{II})}(\bdq,\bdq^{\prime};i\Omega_{n}),\label{eq:DelPhi-inter}
\end{align}
where
\begin{align}
  V_{\nu_{1}\nu_{2}\nu_{3}\nu_{4}}(\bdq,\bdq^{\prime})
  &=\sum_{l_{5},l_{6},l_{7},l_{8}=A,B}
  V_{l_{5}l_{6}l_{7}l_{8}}(\bdq,\bdq^{\prime})
  (U_{\bdq})_{l_{5}\nu_{1}}(U_{\bdq})_{l_{6}\nu_{2}}
  (U_{\bdq^{\prime}})_{l_{7}\nu_{3}}(U_{\bdq^{\prime}})_{l_{8}\nu_{4}},\label{eq:Vq-band}\\
  \Delta G_{\nu_{1}\nu_{2}\nu_{3}\nu_{4}}^{(\textrm{II})}(\bdq,\bdq^{\prime};i\Omega_{n})
  &=T^{2}\sum_{m,m^{\prime}}
  G_{\nu_{1}}(\bdq,i\Omega_{m})G_{\nu_{2}}(\bdq,i\Omega_{n+m})
  G_{\nu_{3}}(\bdq^{\prime},i\Omega_{n+m^{\prime}})G_{\nu_{4}}(\bdq^{\prime},i\Omega_{m^{\prime}}).
  \label{eq:G^II-int-start}
\end{align}
Since $v_{\nu_{1}\nu_{2}}^{x}(\bdq)$ and $e_{\nu_{3}\nu_{4}}^{x}(\bdq^{\prime})$
are odd functions in term of $q_{x}$ and $q_{x}^{\prime}$, respectively, 
and $G_{\nu}(\bdq,i\Omega_{m})$'s are even functions,
the finite terms of $V_{\nu_{1}\nu_{2}\nu_{3}\nu_{4}}(\bdq,\bdq^{\prime})$
in Eq. (\ref{eq:DelPhi-inter}),
i.e., the terms which are finite even after
carrying out $\sum_{\bdq,\bdq^{\prime}}$,
come only from
$V_{ABBA}(\bdq,\bdq^{\prime})=V_{BAAB}(\bdq,\bdq^{\prime})=4J_{\bdq-\bdq^{\prime}}$ [Eq. (\ref{eq:Vq})];
because of this property,
we can replace Eq. (\ref{eq:Vq-band}) by
\begin{align}
  V_{\nu_{1}\nu_{2}\nu_{3}\nu_{4}}(\bdq,\bdq^{\prime})
  =\sum_{l=A,B}4J_{\bdq-\bdq^{\prime}}
  (U_{\bdq})_{l\nu_{1}}(U_{\bdq})_{\bar{l}\nu_{2}}
  (U_{\bdq^{\prime}})_{\bar{l}\nu_{3}}(U_{\bdq^{\prime}})_{l\nu_{4}}.\label{eq:Vq-band-next}
\end{align}
Then, as in $G_{\nu\nu^{\prime}}^{(\textrm{II})}(\bdq,i\Omega_{n})$ [Eq. (\ref{eq:G^II0-next})],
we can replace the sums in Eq. (\ref{eq:G^II-int-start}) by the corresponding integrals:
\begin{align}
  \Delta G_{\nu_{1}\nu_{2}\nu_{3}\nu_{4}}^{(\textrm{II})}(\bdq,\bdq^{\prime};i\Omega_{n})
  &=\Bigl[\int_{\textrm{C}}\frac{dz}{2\pi i}n(z)
    G_{\nu_{1}}(\bdq,z)G_{\nu_{2}}(\bdq,z+i\Omega_{n})
    +A\Bigr]  
    \Bigl[\int_{\textrm{C}^{\prime}}\frac{dz^{\prime}}{2\pi i}n(z^{\prime})
    G_{\nu_{3}}(\bdq^{\prime},z^{\prime}+i\Omega_{n})G_{\nu_{4}}(\bdq^{\prime},z^{\prime})
    +A^{\prime}\Bigr]\notag\\
  &=G_{\nu_{2}\nu_{1}}^{(\textrm{II})}(\bdq,i\Omega_{n})
  G_{\nu_{3}\nu_{4}}^{(\textrm{II})}(\bdq^{\prime},i\Omega_{n}),\label{eq:G^II-int-next-next}
\end{align}
where
$A=T[G_{\nu_{1}}(\bdq,0)G_{\nu_{2}}(\bdq,i\Omega_{n})+G_{\nu_{1}}(\bdq,-i\Omega_{n})G_{\nu_{2}}(\bdq,0)]$,
$A^{\prime}=T[G_{\nu_{3}}(\bdq^{\prime},i\Omega_{n})G_{\nu_{4}}(\bdq^{\prime},0)+G_{\nu_{3}}(\bdq^{\prime},0)G_{\nu_{4}}(\bdq^{\prime},-i\Omega_{n})]$, and
$C$ or $C^{\prime}$
is one of the contours shown in Fig. \ref{figS1}.
By substituting Eq. (\ref{eq:G^II0-next-next}) into Eq. (\ref{eq:G^II-int-next-next})
and performing the analytic continuation $i\Omega_{n}\rightarrow \omega+i\delta$ ($\delta=0+$),
we have
\begin{align}
  \hspace{-10pt}
  \Delta\Phi_{12}^{\textrm{R}}(\omega)
  &=\Delta\Phi_{12}(i\Omega_{n}\rightarrow \omega+i\delta)\notag\\
  \hspace{-10pt}
  =&-\frac{1}{N^{2}}\sum_{\bdq,\bdq^{\prime}}\sum_{\nu_{1},\nu_{2},\nu_{3},\nu_{4}=\alpha,\beta}
  v_{\nu_{1}\nu_{2}}^{x}(\bdq)e_{\nu_{3}\nu_{4}}^{x}(\bdq^{\prime})
  V_{\nu_{1}\nu_{2}\nu_{3}\nu_{4}}(\bdq,\bdq^{\prime})\notag\\
  \hspace{-10pt}
  &\times 
  \int_{-\infty}^{\infty}\frac{dz}{2\pi i}n(z)
  \Bigl\{
  [G_{\nu_{1}}^{\textrm{R}}(\bdq,z)-G_{\nu_{1}}^{\textrm{A}}(\bdq,z)]
  G_{\nu_{2}}^{\textrm{R}}(\bdq,z+\omega)
  +G_{\nu_{1}}^{\textrm{A}}(\bdq,z-\omega)
  [G_{\nu_{2}}^{\textrm{R}}(\bdq,z)-G_{\nu_{2}}^{\textrm{A}}(\bdq,z)]
  \Bigr\}\notag\\
  \hspace{-10pt}
  &\times
  \int_{-\infty}^{\infty}\frac{dz^{\prime}}{2\pi i}n(z^{\prime})
  \Bigl\{
  G_{\nu_{3}}^{\textrm{R}}(\bdq^{\prime},z^{\prime}+\omega)
  [G_{\nu_{4}}^{\textrm{R}}(\bdq^{\prime},z^{\prime})-G_{\nu_{4}}^{\textrm{A}}(\bdq^{\prime},z^{\prime})]
  +[G_{\nu_{3}}^{\textrm{R}}(\bdq^{\prime},z^{\prime})-G_{\nu_{3}}^{\textrm{A}}(\bdq^{\prime},z^{\prime})]
  G_{\nu_{4}}^{\textrm{A}}(\bdq^{\prime},z^{\prime}-\omega)
  \Bigr\}.
\end{align}
Then, by performing the calculations similar to the derivation of Eq. (\ref{eq:L12^0-A}),
we obtain
\begin{align}
  L_{12}^{\prime}
  =&\lim_{\omega\rightarrow 0}
  \frac{\Delta\Phi_{12}^{\textrm{R}}(\omega)-\Delta\Phi_{12}^{\textrm{R}}(0)}{i\omega}\notag\\
  =&\frac{1}{4\pi^{2}iN^{2}}\sum_{\bdq,\bdq^{\prime}}\sum_{\nu_{1},\nu_{2},\nu_{3},\nu_{4}=\alpha,\beta}
  v_{\nu_{1}\nu_{2}}^{x}(\bdq)e_{\nu_{3}\nu_{4}}^{x}(\bdq^{\prime})
  V_{\nu_{1}\nu_{2}\nu_{3}\nu_{4}}(\bdq,\bdq^{\prime})
  \Bigl[
    F_{\nu_{1}\nu_{2}}^{(\textrm{I})}(\bdq)F_{\nu_{3}\nu_{4}}^{(\textrm{II})}(\bdq^{\prime})
    +F_{\nu_{1}\nu_{2}}^{(\textrm{II})}(\bdq)F_{\nu_{3}\nu_{4}}^{(\textrm{I})}(\bdq^{\prime})
  \Bigr],\label{eq:L12'-A}
\end{align}
where
\begin{align}
  \hspace{-10pt}
  F_{\nu\nu^{\prime}}^{(\textrm{I})}(\bdq)
  &=-\frac{1}{2}\int_{-\infty}^{\infty}dz \frac{\partial n(z)}{\partial z}
  \Bigl[G_{\nu}^{\textrm{R}}(\bdq,z)G_{\nu^{\prime}}^{\textrm{R}}(\bdq,z)
    +G_{\nu}^{\textrm{A}}(\bdq,z)G_{\nu^{\prime}}^{\textrm{A}}(\bdq,z)
    -2G_{\nu}^{\textrm{A}}(\bdq,z)G_{\nu^{\prime}}^{\textrm{R}}(\bdq,z)
    \Bigr]\notag\\
  &=2\int_{-\infty}^{\infty}dz \frac{\partial n(z)}{\partial z}
  \textrm{Im}G_{\nu}^{\textrm{R}}(\bdq,z)\textrm{Im}G_{\nu^{\prime}}^{\textrm{R}}(\bdq,z),\label{eq:F^I-last}\\
  \hspace{-10pt}
  F_{\nu\nu^{\prime}}^{(\textrm{II})}(\bdq^{\prime})
  &=\int_{-\infty}^{\infty}dz^{\prime} n(z^{\prime})
  \Bigl[G_{\nu}^{\textrm{R}}(\bdq^{\prime},z^{\prime})G_{\nu^{\prime}}^{\textrm{R}}(\bdq^{\prime},z^{\prime})
    -G_{\nu}^{\textrm{A}}(\bdq^{\prime},z^{\prime})G_{\nu^{\prime}}^{\textrm{A}}(\bdq^{\prime},z^{\prime})
    \Bigr]\notag\\
  &=2i\int_{-\infty}^{\infty}dz^{\prime} n(z^{\prime})
  \Bigl[
    \textrm{Re}G_{\nu}^{\textrm{R}}(\bdq^{\prime},z^{\prime})
    \textrm{Im}G_{\nu^{\prime}}^{\textrm{R}}(\bdq^{\prime},z^{\prime})
    +\textrm{Im}G_{\nu}^{\textrm{R}}(\bdq^{\prime},z^{\prime})
    \textrm{Re}G_{\nu^{\prime}}^{\textrm{R}}(\bdq^{\prime},z^{\prime})
    \Bigr].\label{eq:F^II-last}
\end{align}
A combination of Eqs. (\ref{eq:F^I-last}), (\ref{eq:F^II-last}), and (\ref{eq:L12'-A})
gives Eq. (\ref{eq:L12'}).
\end{widetext}

\section{Derivations of Eqs. (\ref{eq:L12^0-approx}), (\ref{eq:L12^0-approx-band}),
  (\ref{eq:L12'-approx}){--}(\ref{eq:L12_SD})}

We explain the details of the derivations of
Eqs. (\ref{eq:L12^0-approx}), (\ref{eq:L12^0-approx-band}),
(\ref{eq:L12'-approx}){--}(\ref{eq:L12_SD}).
These equations are obtained by deriving the expressions of $L_{12}^{0}$ and $L_{12}^{\prime}$
in the limit $\tau\rightarrow\infty$,
where $\tau=(2\gamma)^{-1}$ is the magnon lifetime. 

First, we derive Eqs. (\ref{eq:L12^0-approx}) and (\ref{eq:L12^0-approx-band}).
Using Eq. (\ref{eq:G^R-band}), we have
\begin{align}
  \textrm{Im}G_{\alpha}^{\textrm{R}}(\bdq,z)
  &=-\frac{\gamma}{[z-\epsilon_{\alpha}(\bdq)]^{2}+\gamma^{2}},\\
  \textrm{Im}G_{\beta}^{\textrm{R}}(\bdq,z)
  &=\frac{\gamma}{[z+\epsilon_{\beta}(\bdq)]^{2}+\gamma^{2}}.
\end{align}
Since $\tau\rightarrow\infty$ corresponds to $\gamma\rightarrow 0$,
we can express
$I_{\nu\nu^{\prime}}^{(\textrm{I})}(\bdq)$ [i.e., Eq. (\ref{eq:I^I})]
in this limit as follows:
\begin{align}
  I_{\alpha\alpha}^{(\textrm{I})}(\bdq)
  &\sim
  \frac{\partial n[\epsilon_{\alpha}(\bdq)]}{\partial \epsilon_{\alpha}(\bdq)}
  \int_{-\infty}^{\infty}dz
  \frac{\gamma^{2}}{\{[z-\epsilon_{\alpha}(\bdq)]^{2}+\gamma^{2}\}^{2}}\notag\\  
  &=
  \frac{\pi}{2\gamma}
  \frac{\partial n[\epsilon_{\alpha}(\bdq)]}{\partial \epsilon_{\alpha}(\bdq)},\label{eq:Iaa^I}\\
  I_{\beta\beta}^{(\textrm{I})}(\bdq)
  &\sim
  \frac{\pi}{2\gamma}
  \frac{\partial n[\epsilon_{\beta}(\bdq)]}{\partial \epsilon_{\beta}(\bdq)},\label{eq:Ibb^I}\\
  I_{\alpha\beta}^{(\textrm{I})}(\bdq)&=I_{\beta\alpha}^{(\textrm{I})}(\bdq)\sim 0.\label{eq:Iab^I}
\end{align}
Combining these equations with Eq. (\ref{eq:L12^0}),
we have
\begin{align}
  L_{12}^{0}&\sim L_{12\alpha}^{0}+L_{12\beta}^{0},\\
  L_{12\nu}^{0}&=\frac{1}{N}\sum_{\bdq}v_{\nu\nu}^{x}(\bdq)e_{\nu\nu}^{x}(\bdq)
  \frac{\partial n[\epsilon_{\nu}(\bdq)]}{\partial \epsilon_{\nu}(\bdq)}\tau.\label{eq:L12^0-nu}
\end{align}
These are Eqs. (\ref{eq:L12^0-approx}) and (\ref{eq:L12^0-approx-band}).

Next, we derive Eqs. (\ref{eq:L12'-approx}){--}(\ref{eq:L12_SD}).
Since $L_{12}^{\prime}$ is given by Eq. (\ref{eq:L12'}),
the remaining task is to derive the expression of $I_{\nu\nu^{\prime}}^{(\textrm{II})}(\bdq)$
in the limit $\tau\rightarrow \infty$.
By performing the similar calculations to
the derivations of Eqs. (\ref{eq:Iaa^I}){--}(\ref{eq:Iab^I}),
we obtain
\begin{widetext}
\begin{align}
  \int_{-\infty}^{\infty}dz n(z)
  \textrm{Re}G_{\alpha}^{\textrm{R}}(\bdq,z)\textrm{Im}G_{\alpha}^{\textrm{R}}(\bdq,z)
  &=-\gamma \int_{-\infty}^{\infty}dz n(z)
  \frac{z-\epsilon_{\alpha}(\bdq)}{\{[z-\epsilon_{\alpha}(\bdq)]^{2}+\gamma^{2}\}^{2}}\notag\\
  &= -\gamma \int_{-\infty}^{\infty}dz n(z)
  \frac{\partial}{\partial z}
  \Bigl\{-\frac{1}{2}\frac{1}{[z-\epsilon_{\alpha}(\bdq)]^{2}+\gamma^{2}}\Bigr\}
  \sim -\frac{\pi}{2}
  \frac{\partial n[\epsilon_{\alpha}(\bdq)]}{\partial \epsilon_{\alpha}(\bdq)}
  ,\label{eq:Faa}\\       
  \int_{-\infty}^{\infty}dz n(z)
  \textrm{Re}G_{\alpha}^{\textrm{R}}(\bdq,z)\textrm{Im}G_{\beta}^{\textrm{R}}(\bdq,z)
  &=\gamma 
  \int_{-\infty}^{\infty}dz n(z)
  \frac{z-\epsilon_{\alpha}(\bdq)}
       {\{[z-\epsilon_{\alpha}(\bdq)]^{2}+\gamma^{2}\}
        \{[z+\epsilon_{\beta}(\bdq)]^{2}+\gamma^{2}\}}
  \sim -\pi
  \frac{n[-\epsilon_{\beta}(\bdq)]}{\epsilon_{\alpha}(\bdq)+\epsilon_{\beta}(\bdq)},\label{eq:Fab}\\
  \int_{-\infty}^{\infty}dz n(z)
  \textrm{Re}G_{\beta}^{\textrm{R}}(\bdq,z)\textrm{Im}G_{\alpha}^{\textrm{R}}(\bdq,z)
  &=\gamma 
  \int_{-\infty}^{\infty}dz n(z)
  \frac{z+\epsilon_{\beta}(\bdq)}
       {\{[z+\epsilon_{\beta}(\bdq)]^{2}+\gamma^{2}\}
        \{[z-\epsilon_{\alpha}(\bdq)]^{2}+\gamma^{2}\}}
  \sim \pi
  \frac{n[\epsilon_{\alpha}(\bdq)]}{\epsilon_{\alpha}(\bdq)+\epsilon_{\beta}(\bdq)},\label{eq:Fba}\\
  \int_{-\infty}^{\infty}dz n(z)
  \textrm{Re}G_{\beta}^{\textrm{R}}(\bdq,z)\textrm{Im}G_{\beta}^{\textrm{R}}(\bdq,z)
  &= -\gamma \int_{-\infty}^{\infty}dz n(z)
  \frac{z+\epsilon_{\beta}(\bdq)}{\{[z+\epsilon_{\beta}(\bdq)]^{2}+\gamma^{2}\}^{2}}\notag\\
  &= -\gamma \int_{-\infty}^{\infty}dz n(z)
  \frac{\partial}{\partial z}
  \Bigl\{-\frac{1}{2}\frac{1}{[z+\epsilon_{\beta}(\bdq)]^{2}+\gamma^{2}}\Bigr\}
  \sim -\frac{\pi}{2}
  \frac{\partial n[\epsilon_{\beta}(\bdq)]}{\partial \epsilon_{\beta}(\bdq)}
  .\label{eq:Fbb}    
\end{align}
\end{widetext}
By combining these equations with Eq. (\ref{eq:I^II}),
we can express $I_{\nu\nu^{\prime}}^{(\textrm{II})}(\bdq)$
in the limit $\tau\rightarrow \infty$ as follows:
\begin{align}
  I_{\alpha\alpha}^{(\textrm{II})}(\bdq)
  &\sim -\pi
  \frac{\partial n[\epsilon_{\alpha}(\bdq)]}{\partial \epsilon_{\alpha}(\bdq)},\\
  I_{\beta\beta}^{(\textrm{II})}(\bdq)
  &\sim -\pi
  \frac{\partial n[\epsilon_{\beta}(\bdq)]}{\partial \epsilon_{\beta}(\bdq)},\\
  I_{\alpha\beta}^{(\textrm{II})}(\bdq)
  &=I_{\beta\alpha}^{(\textrm{II})}(\bdq)
  \sim \pi
  \frac{n[\epsilon_{\alpha}(\bdq)]-n[-\epsilon_{\beta}(\bdq)]}
       {\epsilon_{\alpha}(\bdq)+\epsilon_{\beta}(\bdq)}.
\end{align}
Substituting these equations and Eqs. (\ref{eq:Iaa^I}){--}(\ref{eq:Iab^I})
into Eq. (\ref{eq:L12'}),
we obtain
\begin{align}
  L_{12}^{\prime}\sim
  L_{12\textrm{-intra}}^{\prime}+L_{12\textrm{-inter}1}^{\prime}+L_{12\textrm{-inter}2}^{\prime},
  \label{eq:L12'-approx-A}
\end{align}
where
\begin{align}
  L_{12\textrm{-intra}}^{\prime}
  =&\sum_{\nu=\alpha,\beta}L_{12\textrm{-intra-}\nu}^{\prime},\label{eq:L12'-intra-A1}\\
  L_{12\textrm{-intra-}\nu}^{\prime}
  =&-\frac{2}{N^{2}}\sum_{\bdq,\bdq^{\prime}}
  v_{\nu\nu}^{x}(\bdq)e_{\nu\nu}^{x}(\bdq^{\prime})\tau
  V_{\nu\nu\nu\nu}(\bdq,\bdq^{\prime})\notag\\
  &\times \frac{\partial n[\epsilon_{\nu}(\bdq)]}{\partial \epsilon_{\nu}(\bdq)}
  \frac{\partial n[\epsilon_{\nu}(\bdq^{\prime})]}{\partial \epsilon_{\nu}(\bdq^{\prime})},
  \label{eq:L12'-intra-A2}\\
  L_{12\textrm{-inter}1}^{\prime}
  =&\sum_{\nu=\alpha,\beta}\Bigl\{-\frac{2}{N^{2}}\sum_{\bdq,\bdq^{\prime}}
  v_{\nu\nu}^{x}(\bdq)e_{\bar{\nu}\bar{\nu}}^{x}(\bdq^{\prime})\tau
  V_{\nu\nu\bar{\nu}\bar{\nu}}(\bdq,\bdq^{\prime})\notag\\
  &\times 
  \frac{\partial n[\epsilon_{\nu}(\bdq)]}{\partial \epsilon_{\nu}(\bdq)}
  \frac{\partial n[\epsilon_{\bar{\nu}}(\bdq^{\prime})]}{\partial \epsilon_{\bar{\nu}}(\bdq^{\prime})}
  \Bigr\},
  \label{eq:L12'-inter1-A}
\end{align}and
\begin{align}
  L_{12\textrm{-inter}2}^{\prime}
  =&\sum_{\nu=\alpha,\beta}(L_{\textrm{E}\nu}^{\prime}+L_{\textrm{S}\nu}^{\prime}),
  \label{eq:L12'-inter2-A}\\
  L_{\textrm{E}\nu}^{\prime}
  &=
  \frac{2}{N^{2}}\sum_{\bdq,\bdq^{\prime}}
  v_{\nu\nu}^{x}(\bdq)e_{\alpha\beta}^{x}(\bdq^{\prime})
  V_{\nu\nu\alpha\beta}(\bdq,\bdq^{\prime})\tau\notag\\
  &\times 
  \frac{\partial n[\epsilon_{\nu}(\bdq)]}{\partial \epsilon_{\nu}(\bdq)}
  \frac{n[\epsilon_{\alpha}(\bdq^{\prime})]-n[-\epsilon_{\beta}(\bdq^{\prime})]}
       {\epsilon_{\alpha}(\bdq^{\prime})+\epsilon_{\beta}(\bdq^{\prime})},\label{eq:L12E'-A}\\
  L_{\textrm{S}\nu}^{\prime}
  &=
  \frac{2}{N^{2}}\sum_{\bdq,\bdq^{\prime}}
  v_{\alpha\beta}^{x}(\bdq)e_{\nu\nu}^{x}(\bdq^{\prime})
  V_{\alpha\beta\nu\nu}(\bdq,\bdq^{\prime})\tau\notag\\
  &\times 
  \frac{n[\epsilon_{\alpha}(\bdq)]-n[-\epsilon_{\beta}(\bdq)]}
       {\epsilon_{\alpha}(\bdq)+\epsilon_{\beta}(\bdq)}
  \frac{\partial n[\epsilon_{\nu}(\bdq^{\prime})]}
       {\partial \epsilon_{\nu}(\bdq^{\prime})}.\label{eq:L12S'-A}    
\end{align}
In Eq. (\ref{eq:L12'-inter1-A}),
$\bar{\nu}=\beta$ or $\alpha$ for $\nu=\alpha$ or $\beta$, respectively.
Equations (\ref{eq:L12'-approx-A}){--}(\ref{eq:L12S'-A}) are
Eqs. (\ref{eq:L12'-approx}){--}(\ref{eq:L12_SD}).

\section{Remark on the numerical calculation}

To calculate $L_{\mu\eta}^{0}$ and $L_{\mu\eta}^{\prime}$ numerically,
we perform the momentum summations 
using a $N_{q}$-point mesh of the first Brillouin zone.
Since the sublattice of our ferrimagnetic insulator is described by
a set of primitive vectors, $\bda_{1}={}^{t}(1\ 0\ 0)$,
$\bda_{2}={}^{t}(0\ 1\ 0)$, and $\bda_{3}={}^{t}(0\ 0\ 1)$,
the primitive vectors for the reciprocal lattice are
$\bdb_{1}={}^{t}(2\pi\ 0\ 0)$,
$\bdb_{2}={}^{t}(0\ 2\pi\ 0)$, and $\bdb_{3}={}^{t}(0\ 0\ 2\pi)$.
Thus, in the periodic boundary condition,
momentum $\bdq$ is written in the form
\begin{align}
  \bdq=\frac{m_{x}}{N_{x}}\bdb_{1}+\frac{m_{y}}{N_{y}}\bdb_{2}+\frac{m_{z}}{N_{z}}\bdb_{3},
\end{align}
where $0\leq m_{x}<N_{x}$, $0\leq m_{y}<N_{y}$, and $0\leq m_{z}<N_{z}$
with $N_{x}N_{y}N_{z}=N_{q}=N/2$.  
As a result,
the first Brillouin zone is divided into the $(N_{x}N_{y}N_{z})$-point mesh.
In the numerical calculation
we set $N_{x}=N_{y}=N_{z}=24$ (i.e., $N_{q}=24^{3}$).  

\section{Numerical results at $h=0.08J$ and $0.16J$}

\begin{figure*}
  \includegraphics[width=134mm]{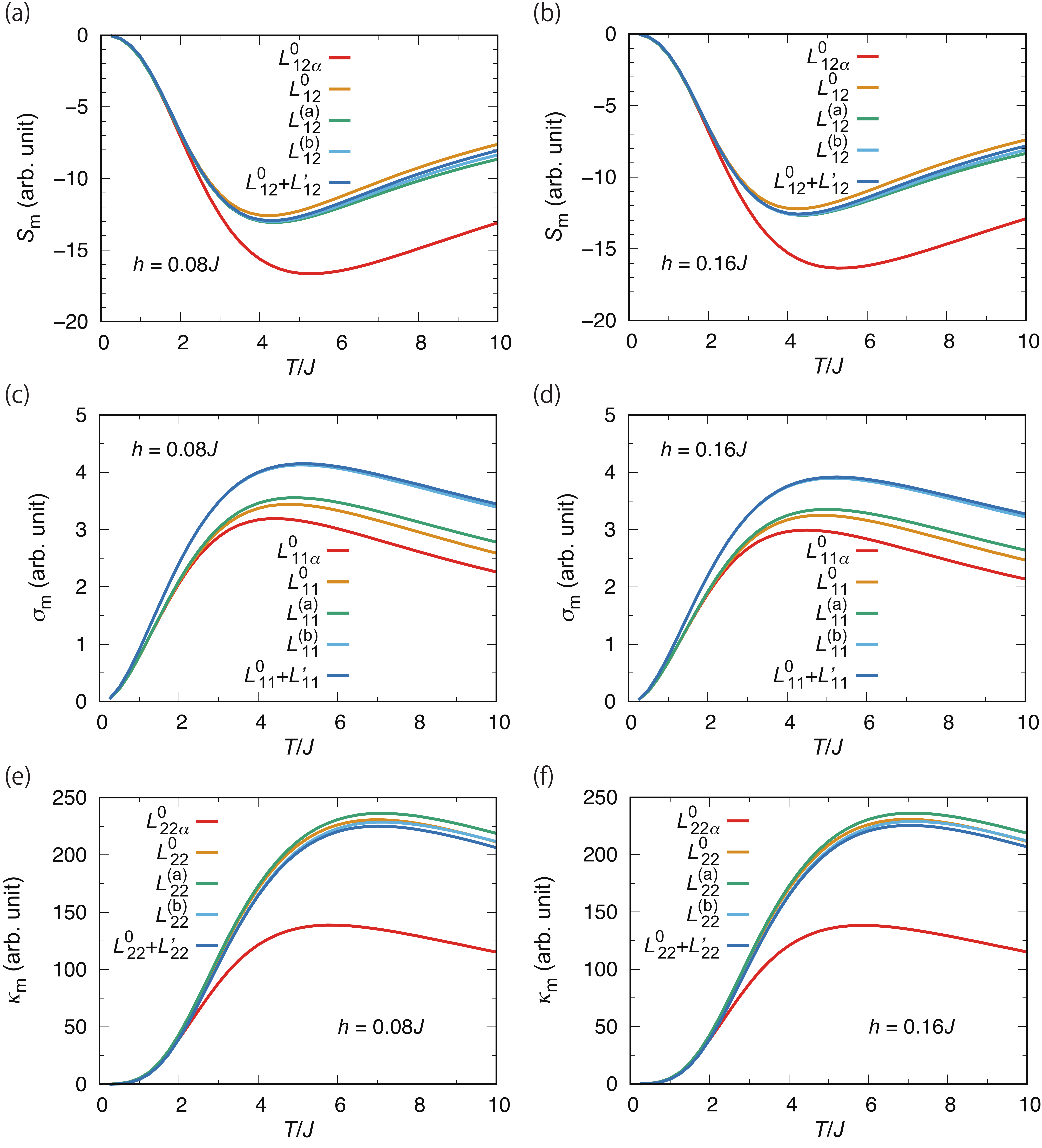}
  \caption{\label{figS2}
    The temperature dependences of
    $S_{\textrm{m}}(=L_{12})$, $\sigma_{\textrm{m}}(=L_{11})$,
    and $\kappa_{\textrm{m}}(=L_{22})$
    at $h=0.08J$ and $0.16J$. $h$ is $0.08J$ in panels (a), (c), and (e)
    and $0.16J$ in panels (b), (d), and (f).
    $L_{\mu\eta}^{(\textrm{a})}$ and $L_{\mu\eta}^{(\textrm{b})}$ 
    are defined as 
    $L_{\mu\eta}^{(\textrm{a})}=L_{\mu\eta}^{0}+L_{\mu\eta\textrm{-intra}}^{\prime}$
    and
    $L_{\mu\eta}^{(\textrm{b})}=L_{\mu\eta}^{0}+L_{\mu\eta\textrm{-intra}}^{\prime}+L_{\mu\eta\textrm{-inter}2}^{\prime}$,
    respectively.
    Note that $L_{\mu\eta}^{0}=L_{\mu\eta\alpha}^{0}+L_{\mu\eta\beta}^{0}$
    and
    $L_{\mu\eta}^{\prime}=L_{\mu\eta\textrm{-intra}}^{\prime}+L_{\mu\eta\textrm{-inter}1}^{\prime}+L_{\mu\eta\textrm{-inter}2}^{\prime}$.
  }
\end{figure*}

We present the additional results of the numerical calculations, 
the temperature dependences of
$S_{\textrm{m}}$, $\sigma_{\textrm{m}}$, and $\kappa_{\textrm{m}}$
at $h=0.08J$ and $0.16J$.
They are shown in Figs. \ref{figS2}(a){--}\ref{figS2}(f).
Comparing these figures with Fig. \ref{fig2},
we see
the results obtained at $h=0.08J$ and $0.16J$ are similar to those obtained at $h=0.02J$.
Namely,
the properties obtained at $h=0.02J$ remain qualitatively unchanged for other values of $h$.

\end{document}